\documentclass[aps,prd,nofootinbib,twocolumn,groupedaddress,showpacs]{revtex4}

\include{graphicx}

\begin{document}

\title{Binary black-hole evolutions of excision and puncture data}

\author{Ulrich Sperhake}
\email[]{Ulrich.Sperhake@uni-jena.de}
\affiliation{Theoretisch-Physikalisches Institut,
             Friedrich-Schiller-Universit\"at\\
             Max-Wien-Platz 1, 07743 Jena, Germany}

\date{\today}

\begin{abstract}
We present a new numerical code developed for the evolution of
binary black-hole spacetimes using different initial data and
evolution techniques. The code is demonstrated to produce
state-of-the-art simulations of orbiting and inspiralling
black-hole binaries with convergent waveforms. We also present
the first detailed study of the dependence of gravitational
waveforms resulting from three-dimensional evolutions of
different types of initial data. For this purpose we compare
the waveforms generated by head-on collisions of superposed
Kerr-Schild, Misner and Brill-Lindquist data over a wide
range of initial separations.
\end{abstract}


\maketitle

\section{Introduction}

In the course of the last two years,
the research area of gravitational wave physics
has entered a very exciting era. On the experimental side,
the first generation of ground-based
Gravitational Wave detectors, LIGO, GEO600, TAMA300 and VIRGO,
are performing observation runs at, and even beyond design sensitivity
\cite{Abbott2006, Abbott2004,
Acernese2005, Ando2005, Hewitson2005}.
At the same time, the simulation of the most
promising sources of gravitational waves, the inspiral of compact
binary systems, has made enormous progress. While approximate
studies based on the Post-Newtonian approach have been able for some time
to accurately
simulate the earlier stages of inspiralling binary systems
\cite{Blanchet2002, Blanchet1995, Blanchet1995a, Blanchet1996, Damour2004,
Koenigsdoerffer2006}, recent developments
in numerical relativity have made possible the simulation of the highly
relativistic final stages of the inspiral and merger of compact binaries
in the framework of fully nonlinear general relativity.

For a long time such simulations have been
troubled by stability problems which
caused evolutions to terminate after times relatively short compared with
the dynamic timescales of the problems under investigation. It is becoming
increasingly clear now, however,
that these problems have been successfully overcome by a combination of
modified formulations of the Einstein equations \cite{Baumgarte1998,
Shibata1995, Bruhat1962, Pretorius2005},
suitable gauge conditions (see e.\,g.\,\cite{Alcubierre2003b,
Bruegmann2004, Campanelli2006a, Baker2006a})
and improved techniques
for the treatment of the singularities inherent to black-hole spacetimes.

Using such modern techniques, Br\"ugmann et.\,al.\,\cite{Bruegmann2004}
obtained the first
simulation of a complete orbit of a black-hole binary in the framework
of the Baumgarte-Shapiro-Shibata-Nakamura (BSSN) formulation
\cite{Baumgarte1998,Shibata1995}, using puncture data \cite{Brandt1997}
and corotating coordinates. More recently,
their results have been confirmed by an improved study
\cite{Diener2006}.
The first waveforms generated in the inspiral and coalescence of black
holes
have been presented by Pretorius \cite{Pretorius2005a, Pretorius2006}
who uses a generalized
harmonic formulation of the Einstein equations combined with special
numerical techniques such as black-hole excision, spatial
compactification and implicit finite differencing.
The latest development, simultaneously discovered by Campanelli et.\,al. and
Baker et.\,al.\,
\cite{Campanelli2006, Baker2006, Campanelli2006a, Baker2006a}, is based
on the evolution of black-hole data of puncture type using
special gauge conditions accommodating the motion of the punctures
across the computational domain. For this reason these simulations
are commonly referred to as {\em moving punctures}. More recently, this
technique has facilitated the investigation of various aspects of
the binary-black-hole coalescence, such as the radiation of linear momentum
by systems of unequal masses and/or spins
\cite{Herrmann2006, Baker2006b, Gonzalez2007, Herrmann2007, Campanelli2007v2,
Koppitz2007, Gonzalez2007a, Tichy2007, Herrmann2007c}, the impact
on the waveforms and merger dynamics of nonvanishing spins
\cite{Campanelli2006b, Campanelli2007} and analysis of the waveforms
in the framework of
post-Newtonian inspiral and black-hole ring-down \cite{Buonanno2006,
Baker2006c, Berti2007, Pfeiffer2007, Husa2007, Hannam2007}.

As in the case of black holes, simulations lasting for several
orbits have also been obtained for neutron star binaries
by several groups
\cite{Shibata2003, Marronetti2004, Miller2004, Anderson2007}.
In more recent
developments the focus is switching to the refinement of the matter
models, as for example by the inclusion of magneto-hydrodynamic effects
(see e.\,g.\,\cite{Duez2006}).
In our work, however, we focus on black-hole systems,
and will therefore exclusively study vacuum spacetimes.

In spite of the dramatic progress in numerical simulations of
black-hole binaries,
there remain important questions to be
answered, in particular with regard to the use of the resulting waveforms
in the ongoing effort to detect and physically interprete
gravitational-wave signals. In particular it will be important to
establish the accuracy of the numerically calculated
waveforms and the consistency of these results with regard to the
use of different types of binary-black-hole initial data and the
evolution techniques used in the codes. First steps in this direction
have been undertaken with regard to the use of evolution techniques and
separation parameters of a given initial-data type.
In Ref.\,\cite{Alcubierre2004},
the impact of black-hole excision was studied in the case of head-on
collisions of Brill-Lindquist data. The results with and without
excision yielded good agreement in that study.
A comparison between plunge waveforms obtained from moving-puncture evolutions
with those resulting from Lazarus calculations \cite{Baker2002} has been
presented in Ref.\,\cite{Baker2006a}. Furthermore,
the waveforms resulting from inspiralling black holes of puncture
type starting from different separations have been found to show excellent
agreement in Refs.\,\cite{Campanelli2006b, Baker2006b}. A comparison of
waveforms obtained from evolving conceptually different types of initial
data has been presented in \cite{Baker2007a}. This study is inhibited,
however, by the difficulties in starting the simulations from
comparable initial configurations as is demonstrated by the nonvanishing
spin in one of the two data sets considered in this work.

The purpose of this paper is two-fold. First, we present a new numerical
code which has been designed to accommodate different types of initial
data, formulations of the Einstein equations as well as singularity treatment.
We demonstrate that the code is capable of producing state-of-the-art
simulations of inspiralling black-hole binaries and extract convergent
waveforms. Second,
we use the code to further progress in the comparison of different initial
configurations by comparing black-hole head-on collisions obtained from
different types of initial data and using different evolution techniques.
Specifically, we compare the results obtained from superposed
Kerr-Schild data evolved in the framework of black-hole excision and
algebraic gauge conditions with those obtained from evolving
Brill-Lindquist as well as Misner data in the framework of the
moving-puncture method.

This paper is structured as follows. We begin with a detailed presentation
of the code in Sec.\,\ref{sec: numerics}. Next, we benchmark the code
in Sec.\,\ref{sec: orbits}
by simulating the inspiral and merger of an orbiting black-hole binary
comparable to those studied in the recent literature. The comparison
of head-on collisions obtained with Brill-Lindquist, Misner and Kerr-Schild
data is given in Sec.\,\ref{sec: head-on} and we conclude with a
discussion of our findings in Sec.\,\ref{sec: conclusions}. Details
on the exact version of the BSSN equations used for this work, the
analytic solution of a boosted black hole in Kerr-Schild coordinates,
extraction of gravitational waves and the performance of the code are
presented in Appendices \ref{sec: BSSN}-\ref{sec: performance}.
Throughout this work we set $G=c=1$ and
use greek indices for spacetime components $0...3$ and latin indices for
spatial components $1...3$.

\section{Computational framework}
\label{sec: numerics}

The simulations presented in this work have been obtained with the
newly developed {\sc Lean} code. This code has been inspired partly
by the {\sc Maya}\footnote{Throughout this work with the {\sc Maya} code
we refer to the version used in Ref.~\cite{Sperhake2005}
which is not to be confused with the new code of the same name
described in Ref.~\cite{Herrmann2006}.}
code \cite{Shoemaker2003, Sperhake2004, Sperhake2005},
and partly by the most recent developments in the simulation of black-hole
data of puncture type \cite{Campanelli2006, Baker2006}.
It is based on the {\sc Cactus} computational toolkit
\cite{Cactusweb}, used for
parallelization and data input/output. Mesh refinement is provided by
{\sc Carpet} \cite{Schnetter2004, Carpetweb}, puncture initial data
by the {\sc TwoPunctures} thorn \cite{Ansorg2004} and horizon finding
by
{\sc AHFinderDirect} \cite{Thornburg1996, Thornburg2004}.
The code achieves dynamic mesh refinement by steering in accordance
with the black-hole motion the regridding option inherent to
the {\sc Carpet} package. While the {\sc Lean} code has been inspired
by {\sc Maya}, it has been written entirely from scratch and
various new features have been added. These are
fourth-order discretization of the spatial
derivatives, the evolution of the BSSN equations using the $\chi$ version
(see below), additional gauge conditions,
time integration using the fourth-order Runge-Kutta (RK) scheme,
dynamic mesh refinement that allows for multiple refinement components
to follow the black-hole motion and merge into single components and
additional initial-data options including puncture data using
the {\sc TwoPuncture} thorn \cite{Ansorg2004} and Misner data.
Furthermore, the different organization of the code has lead to
about five times faster evolutions and a reduction by about a third
in memory requirements compared with {\sc Maya} for a given configuration.
Details on the code's performance for the orbital simulations
and head-on collisions are provided in App.~\ref{sec: performance}.
The key feature of the code for the comparison presented below
is the incorporation in the framework of mesh refinement
of both, dynamic black-hole excision and the
moving-puncture technique used with enormous success in evolutions
of conformally flat initial data.
These features as well as other aspects of the {\sc Lean} code are
described in more detail in the remainder of this section.

\subsection{Formulation of the Einstein equations}

Most of the numerical work in three spatial dimensions has been performed
inside the framework of the canonical ``3+1'' spacetime decomposition
of Arnowitt, Deser and Misner (ADM) \cite{Arnowitt1962} (see also
\cite{York1979} for a detailed discussion). In the notation of
\cite{York1979}, the geometry is
described in terms of the three-dimensional metric $\gamma_{ij}$ and
the extrinsic curvature $K_{ij}$, as well as four gauge functions
$\alpha$ and $\beta^i$ which represent the coordinate
freedom of general relativity. The Einstein field equations
result in six evolution equations each for $\gamma_{ij}$ and $K_{ij}$
as well as four constraint equations,
namely the Hamiltonian and momentum constraints.
These equations are commonly referred to as the ADM equations.

While these equations have been at the heart of most numerical codes
for a long time, the ensuing stability problems have lead to the use
of various alternative formulations of the Einstein equations, most of them
modifications of the ADM equations. The most popular and successful
of these modified schemes is now known as the BSSN system
\cite{Baumgarte1998, Shibata1995} and has been implemented
in the {\sc Lean} code. While the code also allows evolutions using the
Nagy-Ortiz-Reula (NOR) \cite{Nagy2004}
or the generalized harmonic formulation
\cite{Bruhat1962, Pretorius2005}, we have not yet managed to achieve long-term
stable simulations using these systems.
Therefore, all simulations presented
in this work have been obtained with the BSSN system.

The BSSN formulation results from applying the following
modifications to the original ADM
equations: First, a split of the extrinsic curvature
into a trace-free part $A_{ij}$
and the trace $K$, second, a conformal rescaling of the three-metric
and the extrinsic
curvature and, third, the introduction of contracted Christoffel symbols as
separate variables $\tilde{\Gamma}^i$. One thus arrives at a description of the
spacetime in terms of the variables
\begin{eqnarray}
  & \phi = \frac{1}{12} \ln \gamma, \,\,\, &
       \tilde{\gamma}_{ij} = e^{-4\phi} \gamma_{ij}, \nonumber \\
  & K = \gamma^{ij}K_{ij}, & \tilde{A}_{ij} = e^{-4\phi} \left(
       K_{ij}-\frac{1}{3} \gamma_{ij} K \right), \nonumber \\
  & \tilde{\Gamma^i} = \tilde{\gamma}^{mn}\tilde{\Gamma}^i_{mn},
  \label{eq: BSSN_vars}
\end{eqnarray}
as well as the gauge functions $\alpha$ and $\beta^i$. Here, $\gamma$
denotes $\det \gamma_{ij}$ and the definition of $\phi$ implies
that $\tilde{\gamma}=\det \tilde{\gamma}_{ij}=1$. Alternatively
to this choice of variables the {\sc Lean} code also allows
for evolutions using the variable
\begin{equation}
  \chi = e^{-4\phi}
  \label{eq: BSSN_chi}
\end{equation}
as introduced in Ref.~\cite{Campanelli2006}. The complete
evolution equations for both sets of variables are listed
in Appendix \ref{sec: BSSN}.
We refer to the two resulting systems
given by Eqs.~(\ref{eq: gamma})-(\ref{eq: Gamma}) and
Eqs.~(\ref{eq: gamma}), (\ref{eq: tracek}), (\ref{eq: chi})-(\ref{eq: gamma2})
as the $\phi$ and the $\chi$ version of the BSSN equations respectively
in the remainder of this work.

In addition to evolving the BSSN variables according to either of these
systems,
we enforce after each update of
the variables the condition $\tilde{A}^i{}_i=0$, which is a consequence
of the definition of $\tilde{A}_{ij}$ in Eq.\,(\ref{eq: BSSN_vars}).
We find this step to be crucial for the stability of our simulations.
Other modifications to the BSSN equations have been suggested in the
literature [see e.\,g.\,\cite{Yoneda2002}]. We have experimented with
several of these, but not observed any further improvements of the
performance of the code. In particular we do not find it necessary
to enforce the condition $\det\tilde{\gamma}_{ij} = 1$ or
to replace the variable $\tilde{\Gamma}^i$ in terms of the
Christoffel symbols at any stage of the evolution.

\subsection{Initial data}

One main purpose of this paper is to provide a detailed comparison of
binary-black-hole collisions obtained with different initial-data types.
We now describe the different initial data available inside the
code. Specifically, these are puncture, Misner and superposed Kerr-Schild data.

The starting point for binary-black-hole data of the puncture type
is the Schwarzschild solution in isotropic
coordinates, where the spacetime curvature is captured entirely
within the conformal factor $\psi=e^\phi=(1+\frac{m}{2r})$. In the case
of time symmetry, these conformally flat data have been shown
to generalize to an arbitrary number of black holes
by merely adding the individual quotients in the conformal
factor \cite{Misner1957, Brill1963}
\begin{equation}
  \psi = 1+ \sum_i \frac{m_i}{2|\vec{r}-\vec{r}_i|}, \label{eq: psiBL}
\end{equation}
where the index $i$ labels the individual black holes.
This time-symmetric initial configuration of multiple black holes is
known as Brill-Lindquist data. As a further generalization
of these data, spin and momentum can be incorporated in the form of
a nonvanishing extrinsic curvature \cite{Bowen1980}. Finally,
Brandt and Br\"ugmann \cite{Brandt1997} have transformed this type of
data into a form substantially more convenient for the use in numerical
simulations by applying a compactification to the internal asymptotically
flat regions of the holes (see their paper for existence and uniqueness
of the solutions for the Hamiltonian constraint). These data are commonly
referred to as punctures and have been widely used in numerical simulations.

Inside the {\sc Lean} code, initial data of Brill-Lindquist type are implemented
analytically using Eq.\,(\ref{eq: psiBL}). More general classes of puncture
data are made available via the
{\sc TwoPunctures} thorn of Ansorg et.\,al.\,\cite{Ansorg2004},
which solves the Hamiltonian constraint using spectral methods
combined with transformations to a coordinate system specially
adapted to the structure of the binary-black-hole spacetime
[see \cite{Ansorg2004} for details].

The second class of initial data we study in this work are the
axisymmetric Misner data \cite{Misner1960} which represent a
conformally flat spacetime containing two nonspinning
equal-mass black holes at the moment of time symmetry ($K_{ij}=0$).
In Cartesian coordinates the three-metric $\gamma_{ij}$ for this
configuration can be written as
\begin{equation}
  \gamma_{ij} = \psi_{\rm M}^4 \delta_{ij},
\end{equation}
where the conformal factor is given by
\begin{eqnarray}
  \psi_{\rm M} &=& 1+\sum_{n} \frac{1}{\sinh\,n\mu} \left[
      \frac{1}{\sqrt{x^2+y^2+(z+z_n)^2}} \right. \nonumber \\
               && + \left.\frac{1}{\sqrt{x^2+y^2+(z-z_n)^2}} \right]
      \label{eq: Misner_psi} \label{eq: misner} \\
  z_n &=& \coth\,n\mu,
\end{eqnarray}
and $\mu$ is a free parameter determining the initial separation
of the holes $D/M$, where $M$ is the Arnowitt-Deser-Misner (ADM) mass
of the system.

As an alternative to these two conformally flat
data types, the {\sc Lean} code allows
the use of nonspinning
black-hole binary data based on the Kerr-Schild solution
for a single black hole \cite{Kerr1963, Kerr1965}.
The invariance of the structure of the Kerr-Schild data
under boost transformations
has motivated their use in boosted, superposed form. Even though
these superposed
data do not exactly satisfy the Einstein constraints for finite
separation of the holes, they have been studied extensively in the
literature, both as initial data and in the context of binary-black-hole
evolutions (see, for example,
\cite{Matzner1998, Brandt2000, Marronetti2000b, Marronetti2000,
Bonning2003, Sperhake2005}). Whenever we speak of superposed
Kerr-Schild data in the remainder of this work, we thus refer to
this direct superposition of the data which has been used in
evolutions before. We are currently not aware of any time evolutions
presented in the literature that start with Kerr-Schild data
{\em after} applying a constraint solving procedure.

In order to construct the data,
we follow the approach of Ref.~\cite{Sperhake2005}. The solution of
a single boosted, nonspinning Kerr-Schild hole with mass parameter $m$
and velocity $v^i$ is calculated according to the prescription presented
in App.~\ref{sec: bKS}. The two solutions thus obtained for
black holes at positions ${}^{\rm A}x^i$ and ${}^{\rm B}x^i$
are then superposed according to
\begin{eqnarray}
  {}^{\rm KS}
  \gamma_{ij} &=& {}^{\rm A}\gamma_{ij} + {}^{\rm B}\gamma_{ij} - \delta_{ij},
      \label{eq: KS_gamma}\\
  {}^{\rm KS}
  K^i{}_j &=& {}^{\rm A}K^i{}_j + {}^{\rm B}K^i{}_j, \\
  {}^{\rm KS}
  \beta_i &=& {}^{\rm A}\beta_i + {}^{\rm B}\beta_i, \label{eq: KS_beta} \\
  {}^{\rm KS}
  \alpha &=& \left( {}^{\rm A}\alpha^{-2} + {}^{\rm B} \alpha^{-2}-1 \right)
            ^{-1/2}. \label{eq: KS_alpha}
\end{eqnarray}
We note that with this specific superposition, lapse $\alpha$ and
shift $\beta^i$ obey the close-limit condition, i.\,e.\,they lead
to the lapse and shift of a single Kerr-Schild hole in the limit
of zero separation.

\subsection{Gauge conditions}
\label{sec: gauge}

An
important ingredient in the recent success of numerical simulations of
black-hole binaries has been the implementation of improved gauge
conditions. In terms of the ``3+1'' decomposition, the coordinate
invariance of general relativity is represented by the freedom
to arbitrarily specify the lapse function $\alpha$ and the
shift vector~$\beta^i$. While the particular choice of these functions
leaves unaffected the physical properties of the spacetime, it can
have a dramatic effect on the stability properties of a numerical
simulation.

In the past, the majority of gauge conditions have been designed
with the purpose to drive the system of variables towards a
stationary configuration (see e.\,g.\,\cite{Alcubierre2001b,
Alcubierre2003b, Alcubierre2003c}).
In combination with
the use of comoving coordinates, this approach lead to the
first simulation of a complete binary-black-hole orbit
\cite{Bruegmann2004}. More recent developments, however, have
shown a tendency towards allowing the black holes
to move across the computational domain (see e.\,g.\,\cite{Sperhake2004,
Sperhake2005} for single moving black holes and head-on collisions
and \cite{Pretorius2005,Campanelli2006a,Baker2006a, Pretorius2006}
for orbiting black holes). We have implemented
in the {\sc Lean} code both the use of algebraic gauge conditions
along the lines reported in \cite{Sperhake2004}
as well as live-gauge conditions similar to those presented
in \cite{Campanelli2006a,Baker2006a} for the
evolutions of black holes of the moving-puncture type
(see also \cite{vanMeter2006} for a more detailed numerical study
and \cite{Gundlach2006} for an analytic study of
these types of gauge choices).
Experimentally, we have found variations in these live-gauge conditions
to manifest
themselves most conspicuously in the profile of the variables
$\tilde{\Gamma}^i$ near the punctures. In particular, we have noticed that
steep gradients in these functions resulted in poor convergence properties
of the merger time of the black holes, or, worse, instabilities.
We have found optimal performance of our code in this respect by evolving
the gauge variables according to
\begin{eqnarray}
  \partial_t \alpha &=& \beta^i \partial_i \alpha -2 \alpha K,
      \label{eq: alpha} \\
  \partial_t \beta^i &=& B^i, \\
  \partial_t B^i &=& \partial_t \tilde{\Gamma^i} - \eta B^i.
      \label{eq: B}
\end{eqnarray}

Initially we have experimented with setting $\eta=2$, but observed
an instability in the outermost refinement boundary for
coarser resolutions. We have found the choice $\eta=1$ to cure that
instability while preserving the good convergence properties of the code
and therefore use this value throughout this work. The gauge variables
are initialized by
using zero shift with a precollapsed
lapse $\alpha=e^{-2\phi}=\sqrt{\chi}$.

The gauge conditions (\ref{eq: alpha})--(\ref{eq: B}) do not only provide
stable evolutions, but also facilitate a comparatively simple method
to track the black-hole position. As has been shown in
Ref.~\cite{Campanelli2006} the vanishing of $\chi$ at the puncture
in conjunction with Eq.~(\ref{eq: chi}) implies that
\begin{equation}
  \frac{dx^i}{dt} = -\beta^i. \label{eq: punc_tracking}
\end{equation}
We have implemented this relation via interpolation of the shift vector
at the puncture location and subsequent update of the position using a
second-order Runge-Kutta method. In practice
we find excellent agreement between
the resulting locations of the puncture and the coordinate center of
the apparent horizon as calculated by {\sc AHFinderDirect}
from surface integrals of the global coordinates over the horizon.

In the case of the evolutions of Kerr-Schild data, we have also
experimented with these gauge conditions. So far, however, we have not
managed to obtain long-term stable evolutions in this way. We have therefore
reverted to the approach of using algebraic gauge according to the
procedure described in \cite{Sperhake2005}. That is, we prescribe
analytic trajectories ${}^{\rm A}x^i(t)$, ${}^{\rm B}x^i(t)$
for black holes A and B and calculate the resulting gauge functions
by superposing the analytic gauge of the individual holes.
Following \cite{Sperhake2004} we prescribe the analytic
slicing condition in the form of the densitized lapse $Q$.
We thus obtain
\begin{eqnarray}
  {}^{\rm KS}
  \beta^i &=& {}^{\rm KS}\gamma^{ij} ({}^{\rm A}\beta_j
      + {}^{\rm B}\beta_j), \label{eq: beta_IEF} \\
  {}^{\rm KS} Q &=& {}^{\rm KS}\gamma^{-1/2} \left( {}^{\rm A}\alpha^{-2}
        +{}^{\rm B}\alpha^{-2} - 1 \right)^{-1/2}. \label{eq: Q_IEF}
\end{eqnarray}
Here the quantities denoted with an A or B are the analytic
expressions for the individual black holes
and ${}^{\rm KS}\gamma_{ij}$
is the superposed metric defined in Eq.\,(\ref{eq: KS_gamma}).
In practice, we calculate the lapse from its densitized counterpart
and the determinant of the numerical three-metric via
$\alpha = \gamma_{\rm num}^{1/2}Q$. We emphasize that we use
the densitized lapse only for algebraic slicing,
but work with the unmodified lapse in all simulations using
live-gauge conditions.

The trajectories used to evaluate the positions and velocities for the
gauge functions associated with the individual black holes are obtained
from fifth order polynomials $x^i + v^it + a^it^2/2 + j^it^3/6 + q^it^4/24$
during the earlier stages of the infall of the black holes. In a time
interval $t_1 < t < t_2$ we perform a smooth (up to the fourth derivative)
transition of these polynomials to the static function $x^i=0$.
By virtue of the close-limit property of the superposed gauge
(\ref{eq: beta_IEF}), (\ref{eq: Q_IEF}), we thus obtain a smooth
transition of the gauge to that of a single nonspinning Kerr-Schild hole.

The most difficult part in this procedure is to determine the coefficients
$v^i$, $a^i$, $j^i$, $q^i$ and $t_1$, $t_2$ so that one obtains a stable
simulation. We have only managed to obtain stable evolutions using
gauge trajectories that are close to the coordinate trajectories
of the apparent horizon center, in particular in the late stages
of the infall. In practice the black holes collide
along the $z$ axis and we set the $x$ and $y$ components of all
coefficients to zero. The remaining $z$ components are then
obtained iteratively: The black holes are evolved with some initial guess
for the coefficients (normally those used for the simulation with the
next smaller initial separation). The apparent horizon center is tracked
until this simulation becomes unstable and we adjust the parameters to
make the gauge trajectory agree better with the horizon motion. This
process is repeated until a stable simulation is obtained. Some minor
variations of the parameters are possible while preserving the stability
of the simulation but do not have a significant impact on the resulting
waveforms as is discussed below in Sec.~\ref{sec: KStest}.
The exact parameters used for the Kerr-Schild simulations in this work
are given below in Table \ref{tab: bief_traj}.
Unless specified otherwise, we use the trajectories labelled ``a''
in that table.

\subsection{Black-hole excision}
\label{sec: excision}

Evolutions of puncture-type initial data have been performed in the past
both with and without the use of black-hole excision
(see e.\,g.\,\cite{Alcubierre2001b, Alcubierre2003b, Alcubierre2004,
Bruegmann2004, Diener2006}).
Those without excision have commonly been achieved by factoring out
the irregular part of the conformal factor while evolving only the
regular remainder. It is a remarkable and surprising feature of the
moving-puncture evolutions introduced in \cite{Campanelli2006a, Baker2006a},
that these evolutions have been successful using neither excision,
nor the factoring out of the irregular part of the conformal factor.
Below we will follow the same approach for our puncture/Brill-Lindquist
and Misner evolutions.

In order to evolve Kerr-Schild data, however,
we need to use black-hole excision. In contrast to puncture data, the
spatial slices of the Kerr-Schild data do contain the physical singularity
of the black hole at $r=0$, which needs to be removed from the computational
domain. Inside the {\sc Lean} code we have implemented black-hole excision
using either one-sided derivatives or extrapolation techniques.
So far, we have obtained better stability properties using
extrapolation which is the method of choice for all simulations
presented in this work. This particular excision algorithm has been
described in detail in Refs.~\cite{Shoemaker2003, Sperhake2004,
Sperhake2005}. In the {\sc Lean} code the moving excision has now
been generalized to work with moving refinement components. For this
purpose each black hole has been assigned a particular refinement
level it resides in (the finest level in all simulations presented
in this work). Excision for this black hole is then only performed
on this refinement level and communication to coarser levels
is performed exclusively via the restriction procedure inside {\sc Carpet}.
Special care must be taken in the black-hole excision if the refinement
component has been moved because the integer grid indices $i,j,k$ no
longer correspond to the same coordinate position $x,y,z$ as on previous
time steps. Because the set of excision boundary points is stored
in terms of their indices $i,j,k$, rather than their coordinate positions,
we must recalculate the list of excision boundary points every time
the refinement component moves. This process does not involve changing
any of the BSSN variables, however; it merely corrects the book-keeping
of the excision mask.

With a correct list of excision points available at every time step
we thus apply extrapolation of the BSSN variables via second-order
polynomials during each iteration of the Iterated Crank-Nicholson (ICN) cycle
according to the procedure in Sec.~3 of Ref.~\cite{Shoemaker2003}.
After the completion of the whole time step, the code checks for the
position of the black hole and adjusts the center of the excision region
if necessary. As a minor modification compared with the excision
method of the {\sc Maya} code used in \cite{Shoemaker2003, Sperhake2004,
Sperhake2005}, we use the horizon finder
to track the black-hole motion and move the excision region
accordingly.

So far, we have not succeeded in combining black-hole excision with the
fourth-order discretization of the spatial derivatives. The problems
largely arise from the need to use an excision boundary of thickness $\ge2$
to accommodate the wider fourth-order accurate stencils.
For this reason, we use second-order discretization in space for
all simulations using black-hole excision.

\subsection{Mesh refinement}

A further area of remarkable progress in numerical relativity in recent years
is that of mesh refinement, which is used almost routinely now in various
forms in black-hole simulations.
The need for using mesh refinement or essentially equivalent
techniques based on specially adapted coordinates
such as the ``fish-eye transformation'' \cite{Baker2002a}, arises from the
presence of vastly different length scales in the spacetimes. On the one
hand, a code has to resolve the steep gradients near the black-hole horizon,
typically leading to length scales
comparable with the mass of the hole. On the other hand,
the typical wavelength associated with the ringing of a black hole is
one order of magnitude larger. Furthermore, the calculation of accurate
waveforms makes it necessary to extract waves at sufficiently large radii,
ideally, in the wave zone. This requires the use of computational
grids at least two orders of magnitude larger than the radius of a single
black hole. With current computational power, this can only be achieved
inside the framework of mesh refinement.
Simulations of moving black
holes add the extra requirement of dynamic or adaptive refinement.

In the {\sc Lean} code, mesh refinement is provided by the {\sc Carpet}
package. Dynamic refinement based on
{\sc Carpet} has already been reported in \cite{Sopuerta2006}. Here
we use a refined version of this method. {\sc Carpet} provides a
routine which performs a {\em regridding} operation at regular intervals.
That is, it interpretes a steerable parameter string which contains the
exact specifications of all refinement components in terms of their
corner positions.
Inside the {\sc Lean} code, we control this parameter string
via a separate thorn {\sc RegridInfo} which works as follows.
This thorn creates a map between each refinement component and the black hole it
is tied to (a zero entry meaning the component is not tied to black-hole
motion and remains stationary).
The black-hole motion, in turn,
is monitored, either using the horizon
finder or the puncture tracking method according
to Eq.\,(\ref{eq: punc_tracking}).
The corner positions of the refinement components are adjusted
according to the motion of the black holes. The {\sc RegridInfo} thorn
further performs checks on the internal consistency of the
grid specifications and, if necessary,
expands a component to guarantee that all finer
components are accommodated with a minimum number of grid points
between the refinement boundaries. Similarly, it expands components
once the black-hole position comes too close to a refinement boundary.

Finally, the thorn allows for the merger of previously separate
refinement components. This is triggered by the distance between two
components decreasing below a user-specified threshold value.
Again, the parameter string
used by {\sc Carpet} is updated accordingly and the regridding
completes the dynamic adjustment of the
mesh refinement. We find this technique to work very reliably and
to preserve remarkably well the expected convergence properties of
the code, as will be demonstrated below in Sec.\,\ref{sec: orbits}.

For a given simulation the initial grid consists of two types of
cubic refinement levels, $n$ outer levels centered on the origin
which remain stationary throughout the simulation and
$m$ levels with two components centered around either black hole.
In the remainder of this work we specify the exact setup by
giving the resolution $h$ on the finest level as well as the
radius of the cubes excluding ghost zones required for interprocessor
communication. The grid spacing always
increases by a factor of two from one level to the next coarser refinement
level.
For example,
\begin{eqnarray}
  \{(256,128,74,24,12,6)\times (1.5,0.75),\,\,h=1/48\} \nonumber
\end{eqnarray}
specifies a grid with six fixed outer components of radius
$256$, $128$, $74$, $24$, $12$ and $6$ respectively and two refinement
levels with two components each with radius $1.5$ and $0.75$ centered
around either hole. The resolution is $h=1/48$ on the finest level and
successively increases to $8/3$ on the outermost level. In this work
we will use equatorial as well as octant symmetry which reduces the
number of points by a factor of $2$ or 8 respectively. The grid setups
used for the simulations of this work are summarized in
Table \ref{tab: grids}.

\subsection{Discretization of the BSSN equations}
\label{sec: discretization}

In App.~\ref{sec: BSSN} we have listed explicitly the $\phi$ and
$\chi$ version of the BSSN equations as used in the {\sc Lean} code.
The discretization of the spatial derivatives has been implemented
in the form of second-order as well as fourth-order accurate stencils.
With the exception of the advection derivatives
of the form $\beta^i \partial_i f$ these stencils are centered. Advection
derivatives, on the other hand, are approximated with lop-sided
stencils
\begin{eqnarray}
  \partial_x f &=& \frac{1}{2dx} \left(-f_{i+2di,j,k}
                   + 4f_{i+di,j,k} - 3f_{i,j,k} \right), \\
  \partial_x f &=& \frac{1}{12dx} \left(f_{i+3di,j,k} - 6f_{i+2di,j,k}
                   +18f_{i+di,j,k} \right. \nonumber \\
               && \hspace{1.0cm} \left. -10f_{i,j,k} - 3f_{i-di,j,k}\right),
\end{eqnarray}
respectively for second and fourth-order accurate discretization,
with $di=\mathrm{sgn}(\beta^x)$ and likewise for the $y$ and $z$ direction.

Using these representations for the spatial derivatives, the
partial differential equations for the BSSN variables are
integrated in time using the method of lines. Here the time discretization
\begin{table}
  \caption{Grid setup and numerical schemes used for the simulations
           presented in this work. The resolutions used for the
           convergence studies are $h_1=1/48$, $h_2=1/44$, $h_3=1/40$
           for models R1 and BL2,
           $h_1=1/28$, $h_2=1/24$, $h_3=1/20$ for model KS4 and
           $h_1=1/400$, $h_2=1/360$, $h_3=1/320$ for model M4.
           \label{tab: grids}}
  \begin{ruledtabular}
  \begin{tabular}{l|lc}
  Model       &  Scheme      &  grid   \\
  \hline
  R1          &  ICN$\chi_4$ &  $\{(192,128,74,24,12,6)\times(1.5,0.75),h_i\}$ \\
  \hline
  BL1         &  RK$\chi_4$  &  $\{(256,128,96,32,16)\times(4,2,1),1/48\}$ \\
  BL2         &  RK$\chi_4$  &  $\{(256,128,96,32,16)\times(4,2,1),h_i\}$ \\
  BL3         &  RK$\chi_4$  &  $\{(256,128,96,32,16)\times(4,2,1),1/48\}$ \\
  BL4         &  RK$\chi_4$  &  $\{(256,128,96,32,16)\times(4,2,1),1/48\}$ \\
  ISCO        &  ICN$\phi_4$  &  $\{(256,128,88,24,12,8)\times(2.4,1.2,0.6),1/40\}$ \\
  \hline
  KS1         &  ICN$\phi_2$ &  $\{256,128,96,32,16)\times(4,2),1/24\}$ \\
  KS2         &  ICN$\phi_2$ &  $\{256,128,96,32,16)\times(4,2),1/24\}$ \\
  KS3         &  ICN$\phi_2$ &  $\{256,128,96,32,16)\times(4,2),1/24\}$ \\
  KS4         &  ICN$\phi_2$ &  $\{256,128,96,32,16)\times(4,2),h_i\}$ \\
  \hline
  M1          &  ICN$\chi_4$ &  $\{60,30,\frac{45}{2},\frac{15}{2},\frac{15}{4})\times(\frac{15}{16},\frac{15}{32},\frac{15}{64}),\frac{3}{512}$\} \\
  M2          &  ICN$\chi_4$ &  $\{48,24,18,6,3)\times(\frac{3}{4},\frac{3}{8},\frac{3}{16}),\frac{3}{640}$ \\
  M3          &  ICN$\chi_4$ &  $\{40,20,15,5,\frac{5}{2})\times(\frac{5}{8},\frac{5}{16},\frac{5}{32}),\frac{1}{256}\}$ \\
  M4          &  ICN$\chi_4$ &  $\{32,16,12.8,4,2)\times(\frac{1}{2},\frac{1}{4},\frac{1}{8}),h_i\}$ \\
  \end{tabular}
  \end{ruledtabular}
\end{table}
is performed using either the second-order accurate iterated Crank-Nicholson
(ICN) scheme with two iterations \cite{Teukolsky2000} or standard
fourth-order Runge-Kutta (RK) integration. The exact numerical implementation
of the BSSN equations is thus determined by three parameters, the
time integration scheme, the $\phi$ or $\chi$ version and the order
of the spatial discretization. In the remainder of this work, these
different choices are referred to as $\mathrm{RK}\chi_4$,
$\mathrm{ICN}\phi_2$ and so on. The discretizations used
for the simulations in this work are summarized together with
the grid setups in Table \ref{tab: grids}.

The Berger-Oliger--type mesh refinement provided by {\sc Carpet}
requires communication between the refinement levels, the so-called
prolongation and restriction operation (see e.~g.~\cite{Schnetter2004}).
For fourth (second)-order
discretization of the spatial derivatives
we use sixth (fourth)-order accurate prolongation in space
with a total of nine or three buffer zones respectively for the
RK and ICN time discretization (cf.~Sec.~2.3 of \cite{Schnetter2004}).
Prolongation in time
is always performed using third order accuracy. The necessary infrastructure
required for higher order prolongation in time and, thus, genuine
fourth-order accurate communication between the refinement levels
is not available in the currently used implementation of the mesh refinement.
The fourth-order convergence found for the simulations of puncture and
Brill-Lindquist data below indicates that this does not
represent a problem for the type of simulations under discussion in
this work.

\subsection{Wave extraction}

We extract gravitational waves from our numerical simulations by
calculating the Newman-Penrose scalar $\Psi_4$ using the
electromagnetic decomposition of the Weyl tensor
which is described in Appendix \ref{sec: wave_extraction}.
The spatial derivatives required in this calculation are obtained
using either second or fourth-order accurate stencils chosen in
accordance with the spatial discretization of the BSSN evolution
equations.
The calculation of $\Psi_4$ as well as the extraction of modes
has been tested with the analytic expression
calculated for the Teukolsky wave \cite{Teukolsky1982} for
both the $\ell=2,\,m=0$ and $\ell=2,\,m=2$ wave
[cf.\,also Ref.\,\cite{Fiske2005}]. In both cases,
the evolutions have been carried out using the ICN$\phi_2$ implementation
of the BSSN equations and resulted
in second-order convergence of the waveforms.

Once $\Psi_4$ has been calculated on a sphere of constant extraction
radius, the radiated energy and momenta are obtained from
Eqs.\,(22)--(24) in Ref.\,\cite{Campanelli1999}. In practice we perform
these calculations in a post-processing operation using
\begin{figure}[b]
  \includegraphics[angle=-90,width=250pt]{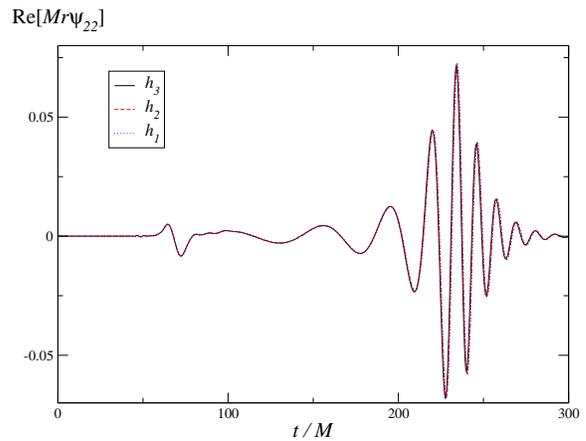}
  \caption{Real part of the $\ell=2$, $m=2$ multipole of $Mr\Psi_4$
           extracted from the $R1$
           simulation at $r_{\rm ex}=60M$ obtained for
           resolutions $h_1$, $h_2$ and $h_3$.}
  \label{fig: wave_orbit}
\end{figure}
the output data of $\Psi_4$. There, we calculate both the total
radiated energy as well as the energy radiated in the dominant modes,
$\ell=2$, $m=\pm 2$ for orbiting configurations and $\ell=2$, $m=0$
for the head-on collisions. For all head-on collisions
we find the dominant
mode to be responsible for $>99\,\%$ of the total radiated
energy; for the inspiral the dominant modes account for about $98.5~\%$
of the total energy.

\section{\label{sec: orbits}Binary black-hole orbits}

Before we compare the head-on collisions of different data types,
we demonstrate the code's capability to produce
evolutions of orbiting black-hole binaries with convergent waveforms. For
this purpose we consider model $R1$ of Table I of Ref.\,\cite{Baker2006a}.
Here two black holes with mass parameter $m=0.483$ start at coordinate
positions $x=\pm 3.257$ with linear momentum parameter $P=\pm 0.133$ in
the $y$ direction.

We evolve this configuration with a setup as specified for model R1
in Table \ref{tab: grids}
using three different resolutions $h_1=1/48$, $h_2=1/44$ and $h_3=1/40$.
The simulations are performed using equatorial symmetry across the
\begin{figure}
  \includegraphics[angle=-90,width=250pt]{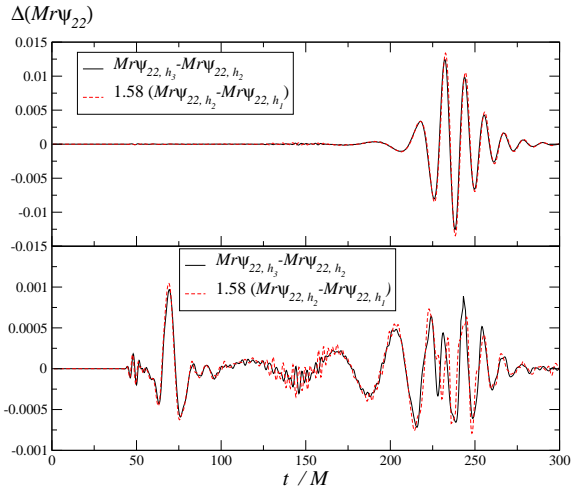}
  \caption{Convergence analysis of the $\ell=2$, $m=2$ multipole
           of $Mr\Psi_4$
           without correcting the phase error (upper panel)
           and after applying a phase correction (lower panel).}
  \label{fig: god1_conv}
\end{figure}
orbital $xy$ plane.

The resulting real part of the
$\ell=2$, $m=2$ mode of the Newman-Penrose scalar
$\Psi_4$ extracted at $r=60\,M$
is shown in Fig.\,\ref{fig: wave_orbit} for all three resolutions.
We first note that the waveforms show good agreement with the results
obtained from similar simulations in the literature \cite{Campanelli2006b,
Baker2006b}. A factor two discrepancy with Fig.\,2 of \cite{Baker2006b}
results from a trivial rescaling depending on the choice of the eigenmode
basis [cf.\,their Eq.\,(4)].

With regard to a convergence analysis,
we first note that the error manifests itself in two forms, a phase
shift and an amplitude difference.
We therefore
study the convergence both with and without applying a phase
correction to align the global maxima of the curves.
For the convergence analysis one commonly assumes
that the
discretization error be dominated by a leading order term
$\mathcal{O}(h^{\alpha})$, so that the numerical solution $f_h$ of
a given grid function is related to the continuum limit $f$ by
$f_h \approx f + Ch^{\alpha}$, where the coefficient C does not depend on the
resolution $h$.
Applying this relation to the three different resolutions
$h_1$, $h_2$, $h_3$ one obtains
\begin{equation}
  \frac{f_{h_3}-f_{h_2}}{f_{h_2}-f_{h_1}} \approx
      \frac{h_3^{\alpha}-h_2^{\alpha}}{h_2^{\alpha}-h_1^{\alpha}}.
\end{equation}
Applied to our case, this relation leads to the value $1.58$
for the case of fourth-order convergence.
The convergence behavior of the $\ell=2$, $m=2$ multipole
of $Mr\Psi_4$ with and without a phase correction
is shown in Fig.\,\ref{fig: god1_conv},
where we have amplified the differences
between the higher resolution runs by the factor $1.58$ expected
for fourth-order convergence. The analysis shows good agreement
\begin{figure}
  \includegraphics[angle=-90,width=250pt]{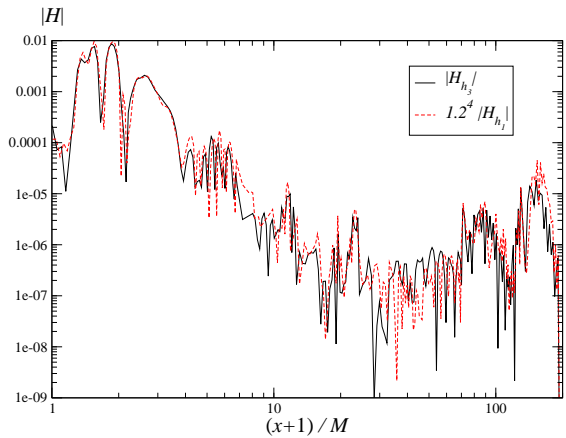}
  \caption{Convergence analysis of the
           Hamiltonian constraint on the $x$ axis at $t=128\,M$,
           shortly after the crossing of the punctures.
           }
  \label{fig: newconv_hc}
\end{figure}
with fourth-order convergence in both cases.

We similarly observe fourth-order convergence for the
total radiated energy extracted at coordinate radii $r_{\rm ex}=50M$,
$60M$ and $70M$. We can use these results to estimate the uncertainties
in the radiated energy resulting from finite resolution and extraction
radii. The standard procedure to assess the impact of the resolution
is to apply Richardson extrapolation, i.~e.~extrapolate the
values obtained for a convergent simulation to the
continuum limit $h\rightarrow 0$. Using this procedure
we obtain $E_{\rm tot}=3.558~\%$, $3.543~\%$ and $3.532~\%$
respectively of the total ADM mass $M$ of the system at
extraction radii $50M$, $60M$ and $70M$. This corresponds to an
estimate of the discretization error in the radiated energy of
about $1~\%$ for the high resolution $h_1=1/48$ and about $2~\%$ for
the low resolution $h_3=1/40$.

In complete analogy to the
procedure used to study the
convergence with grid resolution $h$, we use these values to
estimate the dependency of the radiated energy on the extraction
radius. We find the resulting error to be modeled well by a
$1/r_{\rm ex}$ fall-off, i.~e.
\begin{equation}
  E_{\rm tot} \approx \left. E_{\rm tot}\right|_{r_{\rm ex}=\infty}
                      + \mathcal{O}\left( \frac{1}{r_{\rm ex}} \right).
\end{equation}
Extrapolation of the results obtained at finite extraction radii
thus gives a total radiated energy of $E_{\rm tot} = 3.466~\%$ of the
total ADM mass as well as an estimate for the error arising out of the
use of a finite extraction radius of $2.7~\%$, $2.3~\%$
and $2.0~\%$ respectively for $r_{\rm ex}=50M$, $60M$ and $70M$.

For the simulations presented in this work we universally find
the errors due to finite differencing and finite extraction
radius to point in opposite directions: finite resolution
leads to underestimating the amount of radiated energy, a finite
extraction radius overestimates the energy. We therefore feel
justified in using the sum of the individual errors as a
conservative upper limit for the total error. In this case, we
obtain a numerical error of $3~\%$ for the high resolution
simulation using $r_{\rm ex}=70M$.

Repeating the same calculation without including the
artificial radiation burst due to the initial data\footnote{In practice
we ignore contributions at $t<r_{\rm ex}+30M$ in the waveforms.},
we obtain
a total radiated energy of $E_{\rm rad}=3.408~\%$ of the ADM mass.
We note that this result for the energy shows
excellent agreement with those presented in the literature
(cf.\,Table III of \cite{Baker2006b}).

As a further test of the
code we follow Ref.~\cite{Campanelli2006a} and check the
convergence properties of the Hamiltonian constraint
on an equatorial axis shortly after the crossing of the punctures.
The result is shown in Fig.\,\ref{fig: newconv_hc} where
we have amplified the high resolution result by a factor of $1.2^4$
as expected for fourth-order convergence. In spite of the presence
of some numerical noise, the figure demonstrates compatibility with
overall fourth-order convergence of the simulations.
We believe the larger amount of noise, as compared with
the results of
\cite{Campanelli2006a}, to be a consequence of the use of mesh refinement
and the discontinuous error terms at the refinement boundaries.

Unfortunately we are currently not able to obtain
similar orbital simulations with Kerr-Schild data for want of
suitable live-gauge conditions analogous to
Eqs.\,(\ref{eq: alpha})--(\ref{eq: B}). We therefore perform the
comparison between these two data types and the Misner data
inside the framework of head-on collisions.

\section{Head-on collisions}
\label{sec: head-on}

Head-on collisions represent the simplest form of black-hole binaries
and have been studied numerically
in various forms for a long time. The majority of
such simulations has been performed using data of Misner
\cite{Misner1960} or Brill-Lindquist
type (see e.\,g.\,\cite{Smarr1976, Anninos1993, Anninos1995c, Alcubierre2003b,
Zlochower2005, Fiske2005}). As an alternative, collisions using Kerr-Schild
data have been investigated in \cite{Sperhake2005}. Here we will study in
detail head-on collisions of all three data types
and compare the results.

The time evolutions of these two types of initial data present
different difficulties and therefore require different evolution
techniques. In summary, these are the use of second-order
differencing, algebraic
gauge conditions and black-hole excision for the Kerr-Schild data,
whereas Brill-Lindquist data are evolved using
fourth-order discretization without using the black-hole excision
procedure described in Sec.~\ref{sec: excision}.
As will be demonstrated below,
we have obtained satisfactory convergence performance for the Kerr-Schild
evolutions using the $\phi$ version of the BSSN equations. In the case of
Brill-Lindquist data, we find the $\chi$ version more successful
in providing fourth-order convergence.
The Misner data are conceptually similar to Brill-Lindquist data.
Experimentally, however, we have found the Misner data to lead to substantially
larger amounts of numerical noise originating from the refinement
boundaries when evolved in time with the Runge-Kutta method. Below we
will show that the noise level is acceptable when using the second-order
accurate ICN scheme instead.

In the notation of Sec.~\ref{sec: discretization} we therefore evolve
the Kerr-Schild data using the ICN$\phi_2$ scheme, the Brill-Lindquist data
with the\footnote{with the exception of the simulations
in Figs~\ref{fig: bl_isco} and \ref{fig: new_bl_D}
which are not used in this
quantitative comparison} RK$\chi_4$
and Misner data with the ICN$\chi_4$ scheme. The resulting
accuracy and convergence properties will be studied in detail
in Sec.~\ref{sec: tests}.

\subsection{Choice of initial parameters}

A fundamental difficulty in the comparison between simulations
of Brill-Lindquist, Kerr-Schild and Misner data is the
physical interpretation of the initial-data sets.
We first note that there
exists no general method to rigorously quantify the degree to which two
such initial configurations represent the same physical scenario.
As an approximation, we determine the initial parameters as follows.
\begin{figure}
  \includegraphics[angle=-90,width=250pt]{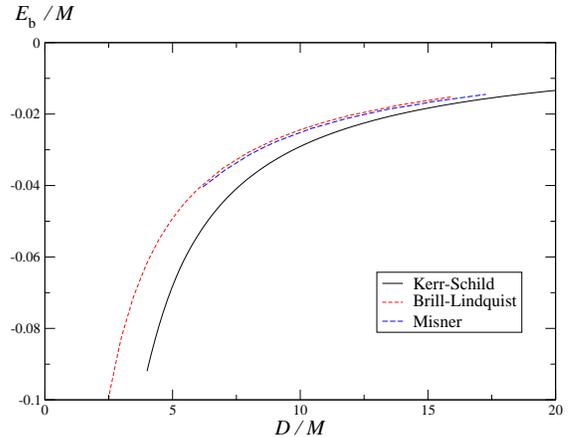}
  \caption{Binding energy $E_b/M$ for Kerr-Schild (solid),
           Brill-Lindquist (short-dashed) and Misner (long-dashed)
           initial-data sets as function of the coordinate
           distance $D/M$ of the holes.}
  \label{fig: eb_BL_KS_Mi}
\end{figure}
First, we start the head-on collisions with two black holes of equal
mass at rest,
and thus eliminate the question of choosing initial linear momenta
and mass ratios.

Except for a rescaling of the entire spacetime
corresponding to a rescaling of the system's total mass,
an initial configuration for a head-on collision of nonspinning,
equal-mass black holes is characterized by one free parameter which
can be viewed as a measure for the initial separation of the black holes
or the binding energy of the system. In the case of Misner data,
this degree of freedom is represented by the parameter $\mu$ in
Eq.~(\ref{eq: Misner_psi}), whereas Brill-Lindquist and Kerr-Schild data
require coordinate positions of the black holes so that the
free parameter is the initial coordinate separation $D$.
For the comparison of the different data types we fix the
remaining free parameter by demanding that
all three versions of the binary-black-hole spacetime
have identical binding energy
\begin{equation}
  \frac{E_{\rm b}}{M} = 1 - \frac{M_1 + M_2}{M},
\end{equation}
where the irreducible black-hole masses $M_1$, $M_2$ are given
by their respective apparent horizon masses. For illustration we
plot in Fig.~\ref{fig: eb_BL_KS_Mi} the binding energy as a function
of the coordinate distance $D$. For Misner data, the initial black-hole
centers are approximated by their apparent horizon position which
is $\pm1.0$ for all simulations discussed in this work.

The freedom in rescaling the spacetimes is fixed by demanding the
total ADM mass to be unity for Kerr-Schild and Brill-Lindquist data which
implies bare mass parameters $m_1=m_2=0.5$ in both cases. In contrast,
the conformal factor for the Misner data in Eq.~\ref{eq: misner} does
not contain a bare mass parameter and the ADM mass depends on the value
of $\mu$. Specifically, it decreases for larger $\mu$ and thus
implies a larger black-hole separation $D/M$. In our simulations we
have taken into account the different ADM masses of Kerr-Schild and
Brill-Lindquist data on the one hand and Misner data on the other
by using numerically higher resolutions for the Misner models.
Relative to the ADM mass, however, the resolutions are rather similar
to those used for Brill-Lindquist data.
The exact ADM masses for all simulations
discussed in this comparison are
given in the third column of Table \ref{tab: models}.

Using this procedure, we have determined four different models with binding
\begin{table*}
  \caption{Summary of the simulations performed in this work. Simulation
           R1 is the inspiral simulation described in Sec.~\ref{sec: orbits}.
           The other simulations are the head-on collisions performed for
           the comparison of
           Brill-Lindquist, Misner and superposed Kerr-Schild data.
           $M$ is the total ADM mass of the spacetime,
           $D$ the initial coordinate separation of the holes
           (for Misner data we list the parameter $\mu$ instead) and
           $E_b$ the binding energy $M-M_{1}-M_{2}$.
           $E_{\rm tot}$,
           $E_{\rm ini}$ and $E_{\rm rad}$ are the total radiated
           energy, the energy contained in the spurious initial burst
           and the energy radiated in the inspiral and merger.
           The uncertainties included are those due to finite
           differencing, finite extraction radius and the
           uncertainties in separating the merger signal
           from the spurious initial burst.
           For the Kerr-Schild data
           we also list the radiated energies obtained from extrapolation to
           $r_{\rm ex}\rightarrow \infty$ with uncertainties due
           to finite differencing and the interference of the initial burst.
           \label{tab: models}}
  \begin{ruledtabular}
  \begin{tabular}{l|ccc|ccc}
  Model       &  $D$ or $\mu$   &  $M$              &  $E_{\rm b}/M$  &  $E_{\rm tot}/M$  &  $E_{\rm ini}/M$  &  $E_{\rm rad}/M$ \\
  \hline
  R1          &  6.5   &  0.996            &  $-0.0145$      &  $(3.466\pm0.104)~\%$     &  $(0.058\pm 0.002)~\%$     &  $(3.408\pm 0.102)~\%$    \\
  \hline
  BL1         &  8.6   &  1                &  $-0.0290$      &  $(0.0553\pm 0.0008)~\%$    &  $(0.0031\pm0.0001)~\%$    &  $(0.0522\pm0.0008)~\%$   \\
  BL2         &  10.2  &  1                &  $-0.0240$      &  $(0.0553\pm 0.0008)~\%$    &  $(0.0022\pm0.0001)~\%$    &  $(0.0531\pm0.0008)~\%$   \\
  BL3         &  12.5  &  1                &  $-0.0197$      &  $(0.0557\pm 0.0008)~\%$    &  $(0.0014\pm0.0001)~\%$    &  $(0.0543\pm0.0008)~\%$   \\
  BL4         &  14.6  &  1                &  $-0.0169$      &  $(0.0564\pm 0.0009)~\%$    &  $(0.0009\pm0.0001)~\%$    &  $(0.0555\pm0.0008)~\%$   \\
  \hline
  KS1         &  10.0  &  1                &  $-0.0290$      &  $(0.1099\pm 0.0175)~\%$    &  $(0.0540\pm0.0119)~\%$    &  $(0.0560\pm0.0123)~\%$  \\
  $r_{\rm ex}\rightarrow \infty$
              &        &                   &                 &  $(0.0963\pm 0.0029)~\%$    &  $(0.0438\pm0.0049)~\%$    &  $(0.0525\pm0.0052)~\%$  \\

  KS2         &  12.0  &  1                &  $-0.0240$      &  $(0.0962\pm 0.0154)~\%$    &  $(0.0325\pm0.0072)~\%$    &  $(0.0617\pm0.0119)~\%$  \\
  $r_{\rm ex}\rightarrow \infty$
              &        &                   &                 &  $(0.0844\pm 0.0025)~\%$    &  $(0.0284\pm0.0019)~\%$    &  $(0.0560\pm0.0032)~\%$  \\

  KS3         &  14.0  &  1                &  $-0.0197$      &  $(0.0888\pm 0.0142)~\%$    &  $(0.0227\pm0.0040)~\%$    &  $(0.0661\pm0.0110)~\%$  \\
  $r_{\rm ex}\rightarrow \infty$
              &        &                   &                 &  $(0.0789\pm 0.0024)~\%$    &  $(0.0198\pm0.0010)~\%$    &  $(0.0591\pm0.0023)~\%$  \\

  KS4         &  16.0  &  1                &  $-0.0169$      &  $(0.0855\pm 0.0137)~\%$    &  $(0.0163\pm0.0028)~\%$    &  $(0.0692\pm0.0112)~\%$  \\
  $r_{\rm ex}\rightarrow \infty$
              &        &                   &                 &  $(0.0751\pm 0.0023)~\%$    &  $(0.0140\pm0.0007)~\%$    &  $(0.0611\pm0.0021)~\%$  \\
  \hline
  M1          &  3.573&  0.231            &  $-0.0290$      &  $(0.0555\pm0.0017)~\%$    &  $(0.0031\pm0.0001)~\%$    &  $(0.0524\pm(0.0016)~\%$   \\
  M2          &  3.757&  0.191            &  $-0.0240$      &  $(0.0556\pm0.0017)~\%$    &  $(0.0021\pm0.0001)~\%$    &  $(0.0535\pm(0.0016)~\%$   \\
  M3          &  3.948&  0.157            &  $-0.0197$      &  $(0.0560\pm0.0017)~\%$    &  $(0.0014\pm0.0001)~\%$    &  $(0.0546\pm(0.0017)~\%$   \\
  M4          &  4.096&  0.135            &  $-0.0169$      &  $(0.0567\pm0.0017)~\%$    &  $(0.0009\pm0.0001)~\%$    &  $(0.0558\pm(0.0017)~\%$   \\
  \end{tabular}
  \end{ruledtabular}
\end{table*}
energies $E_{\rm b}/M = -0.0290$, $-0.0240$, $-0.0197$ and $-0.0169$.
These models are listed in Table \ref{tab: models} as BL$1-4$, KS$1-4$ and
M$1-4$.

\subsection{Testing the code}
\label{sec: tests}

In this section we calibrate the performance of the code in the
case of head-on collisions of all data types by performing
convergence tests and investigating other error sources.
For Brill-Lindquist data we also use
results published in Ref.~\cite{Fiske2005}
starting from the approximate separation
of the innermost stable circular orbit (ISCO).

\subsubsection{Brill-Lindquist data}
\label{sec: test_bl}

\begin{figure}
  \includegraphics[angle=-90,width=250pt]{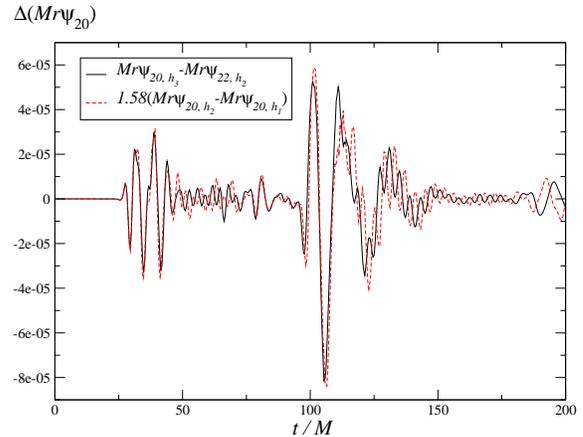}
  \caption{Convergence analysis for the $\ell=2$, $m=0$ mode
           of the Newman-Penrose scalar $\Psi_4$ extracted at
           $r_{\rm ex}=40M$. The difference between the
           runs with higher resolutions has been amplified
           by a factor 1.58 expected for fourth-order convergence.}
  \label{fig: new_conv_BL}
\end{figure}

We assess the discretization error of the Brill-Lindquist simulations
by evolving model BL2 of Table \ref{tab: models} using three
different resolutions $h_1=1/48$, $h_2=1/44$ and $h_3=1/40$
with a constant Courant factor of $dt/dx=1/2$.
We have studied the convergence of the resulting gravitational waveforms
in complete analogy to the procedure used in Sec.~\ref{sec: orbits}
for black-hole inspiral.
We observe fourth-order convergence as is demonstrated in
Fig.~\ref{fig: new_conv_BL} for the $\ell=2$, $m=0$ multipole
extracted at $r_{\rm ex}=40M$. We similarly observe fourth-order
convergence for the total radiated energy and use Richardson
extrapolation to estimate the discretization error. We find the
relative error at a resolution $h=1/44$ to be less than $0.5~\%$ and
use this value as a conservative upper limit.

In order to assess the impact of extracting waves at finite
radii we have studied the wave signal at extraction
radii $40M$, $70M$ and $90M$.
\begin{figure}
  \includegraphics[angle=-90,width=250pt]{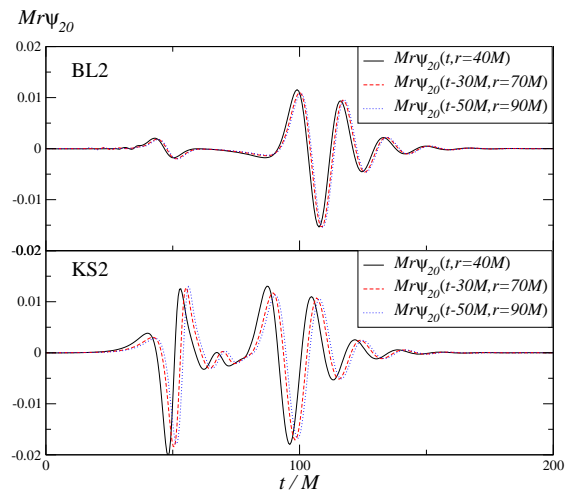}
  \caption{The $\ell=2$, $m=0$ multipole obtained for models
           BL2 (upper) and KS2 (lower panel) extracted at
           $r_{\rm ex}=40M$, $70M$ and $90M$.
           \label{fig: new_rex}}
\end{figure}
The resulting $\ell=2$, $m=0$ multipole of the Newman-Penrose scalar $\Psi_4$
is shown in the upper panel of Fig.~\ref{fig: new_rex}.
We estimate the uncertainty due to the extraction radius in the
same way as in Sec.~\ref{sec: orbits} and find
the relative error in the total radiated energy to be of the order of
$1~\%$ for $r_{\rm ex}=40M$ and less for the larger radii $70M$ and $90M$.

\begin{figure}
  \includegraphics[angle=-90,width=250pt]{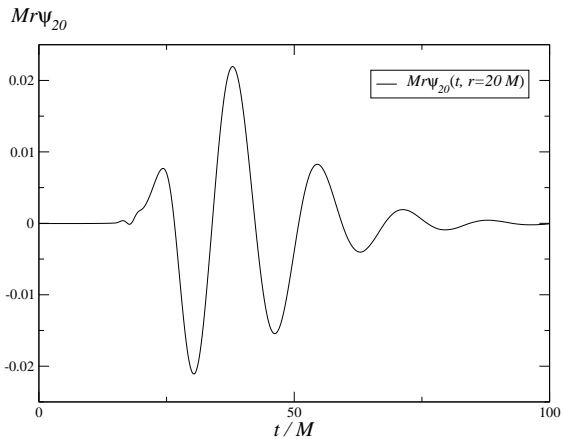}
  \caption{The $\ell=2$, $m=0$ mode of
           $Mr\Psi_4$ extracted at $r_{\rm ex}=20M$ from a
           head-on collision of Brill-Lindquist data starting
           from the approximate ISCO separation (cf.~Fig.~5 in
           Ref.~\cite{Fiske2005}.
           \label{fig: bl_isco}}
\end{figure}

As a further test of our code, we compare the waveforms
obtained for Brill-Lindquist data with those available in the literature.
The head-on collisions presented commonly start with time
symmetric initial data of two holes at positions $\pm1.1515\,M$. This
value has been calculated in Refs.~\cite{Cook1994, Baumgarte2000} for the
ISCO. The
$\ell=2$, $m=0$ mode of the Newman-Penrose
scalar $\Psi_4$ of this configuration has been calculated in
\cite{Fiske2005} at an extraction radius $r_{\rm ex}=20M$. In
Fig.\,\ref{fig: bl_isco} we plot our result extracted at the
same radius and obtained using the setup labelled ``ISCO'' in
Table \ref{tab: grids}. Up to the trivial rescaling factor of two mentioned
above, we find excellent agreement with Fig.~5 in Ref.\,\cite{Fiske2005}.
To our knowledge, similar results obtained with
larger initial separations of the holes have not yet been published. It is part
of the motivation of this work to provide such an extension of the
existing work.

We finally address one conceptual difference between the Brill-Lindquist and
Misner data on the one side and Kerr-Schild data on the other. Whereas the
former initial data are time-symmetric,
the superposed Kerr-Schild data do not satisfy this requirement,
even for vanishing velocity parameter. We thus
cannot rule out that the individual Kerr-Schild holes do actually
have a small boost and thus represent a slightly different
physical configuration.
Unfortunately, there exists no rigorous way to quantify the linear momenta
of the individual holes in the Kerr-Schild spacetime, although the
hypothesis of small momenta is compatible with the small nonzero
initial coordinate velocities of the apparent horizon positions
of the order of $v=0.05$ towards each other
which we observe for the Kerr-Schild holes.
For this reason
we proceed differently and instead consider the impact of small initial
boosts as an additional uncertainty in our study.
We quantitatively study this effect using a modified version of
model BL4. Specifically, we use
puncture data, where
initial linear momenta pointing towards each other
are applied to the individual black holes
in the form of nonzero Bowen-York \cite{Bowen1980}
momentum parameters $P=0.035$ and $P=0.067$.
All other parameters for these
puncture models are kept at the values of model BL4.

\begin{figure}
  \includegraphics[angle=-90,width=250pt]{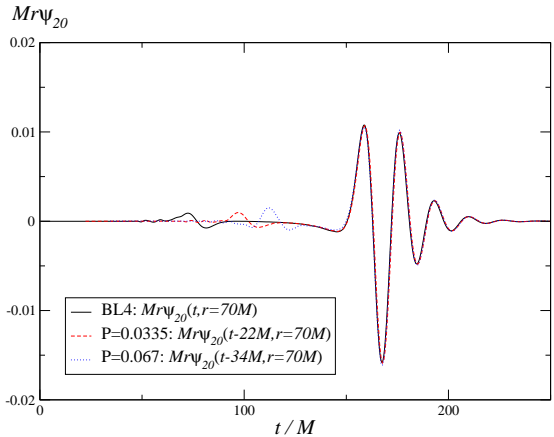}
  \caption{The $\ell=2$, $m=0$ mode of
           $Mr\Psi_4$ extracted at $r_{\rm ex}=70M$
           for model BL4 (solid), as well as puncture
           initial data with $P=\mp0.0335$ (dashed) and
           $P=\mp0.067$ (dotted) for the holes starting
           at coordinate positions $z=\pm 7.3$.
           \label{fig: bl_vel}}
\end{figure}

In Fig.~\ref{fig: bl_vel} we show the resulting waveforms
at $r_{\rm ex}=70M$ shifted in time to align their
global extrema.
The differences in the waveforms are rather small and we obtain for the
energy radiated in the infall and merger $E_{\rm rad}=0.05601\pm0.0008$
and $0.05795\pm0.0009$ respectively of the ADM mass. Compared
with the nonboosted result $0.0555\pm0.0008$, this corresponds to
systematic deviations of about $1~\%$ and $4~\%$.

\subsubsection{Kerr-Schild data}
\label{sec: KStest}

By evolving the Kerr-Schild models of Table \ref{tab: models}
at different resolutions,
we find model KS4 with the largest initial separation to
exhibit the largest uncertainties. We therefore focus in our
convergence analysis on this model to obtain conservative upper limits
for the discretization error.
We have evolved this model using the finest resolutions
$h_1=1/28$, $h_2=1/24$ and $h_3=1/20$ and a constant
Courant factor of $1/4$. Compared with the Brill-Lindquist data
we have to use these seemingly coarser resolutions on the finest
level because the coordinate radius of the apparent horizon
is larger in the Kerr-Schild case and the use of excision
prohibits the use of refinement components inside the apparent
horizon. We emphasize, however, that relative to the coordinate
radius of the horizon, about $r_{\rm ah}=2M_{1,2}$ for Kerr-Schild data
and $r_{\rm ah}=M_{1,2}/2$ for Brill-Lindquist data, our setup results in
a finer resolution in the Kerr-Schild case.

\begin{figure}
  \includegraphics[angle=-90,width=250pt]{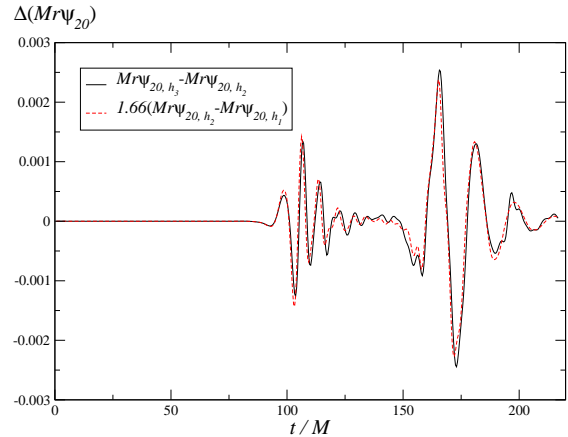}
  \caption{Convergence analysis of the $\ell=2$, $m=0$ mode
           of $\Psi_4$ obtained at $r_{\rm ex}=90M$
           for model KS4 using resolutions
           $h_1=1/28$, $h_2=1/24$ and $h_3=1/20$. The differences
           between the high resolution results has been amplified by
           a factor of $1.66$ expected for second-order convergence.
           \label{fig: new_conv_BIEF}}
\end{figure}

The resulting differences in the $\ell=2$, $m=0$ multipole of the
Newman-Penrose scalar are shown in Fig.~\ref{fig: new_conv_BIEF}
and demonstrate second-order convergence. Using
Richardson extrapolation as before, we obtain an error
of $3~\%$ in the radiated energy due to the discretization of the
Einstein equations.

In comparison with the Brill-Lindquist data, we observe larger
differences in the amplitude of the gravitational waves extracted
at different radii. This is shown in the lower panel of
Fig.~\ref{fig: new_rex} where we plot the $\ell=2$, $m=0$
multipole obtained for model KS2 at $r_{\rm ex}=40M$, $70M$
and $90M$. Systematically investigating the total radiated
energy for all Kerr-Schild simulations we find the results
compatible with a $1/r_{\rm ex}$ fall-off as in Sec.~\ref{sec: orbits}.
For example,
the energies extracted from model KS4 at $40M$, $70M$ and
$90M$ using a resolution $h=1/24$ are $0.0984$, $0.884$
and $0.855~\%$ of the total mass $M$ respectively. Extrapolation
to infinite radius results in $0.0752~\%$ of the
mass and a relative error of $13~\%$ at extraction
radius $r_{\rm ex}=90M$. We find the uncertainties due to the
extraction radius to be very similar for all other Kerr-Schild
simulations and to be essentially independent of the
grid resolution.

In view of this large uncertainty
we will always
present in the remainder of this work two values for the
energies resulting from simulations of Kerr-Schild data.
The numerical values obtained at the largest extraction
radius $r_{\rm ex}=90M$, together with uncertainties
due to extraction radius, discretization and interference
of the initial pulse, and the value extrapolated to
infinite radius with uncertainties due to discretization and
the initial pulse are listed in Table \ref{tab: models}.

A further difficulty in the case of the Kerr-Schild data
arises from the relatively large amount of spurious
radiation due to the initial data. This spurious
radiation manifests itself as a pulse in the
waveform starting at $t\approx r_{\rm ex}$. In the
two panels of Fig.~\ref{fig: new_rex}, for example,
the initial burst leads to local extrema in the waveforms
near $t\approx50M$. The amplitude of the pulse is, however,
substantially larger for the Kerr-Schild (lower panel)
than the Brill-Lindquist (upper panel) data. For the smaller
separations used in our analysis, it becomes nontrivial
to disentangle this pulse from the actual merger signal
and it is not entirely clear, how much radiated energy is
due to the black-hole merger and how much due to the
spurious pulse. We attempt to bracket these uncertainties
by using a variable threshold $t_{\rm thresh}$ so that radiation
at $t<t_{\rm thresh}$ is considered part of the initial pulse
and radiation at larger $t$ part of the merger signal.
At a given extraction radius we then
vary $t_{\rm thresh}$ in the range $r_{\rm ex}+30M$ to
$r_{\rm ex}+40M$. The two resulting energy contributions
are given by their average values obtained over this interval
plus or minus an error given by the upper and lower bounds.

\begin{figure}
  \includegraphics[angle=-90,width=250pt]{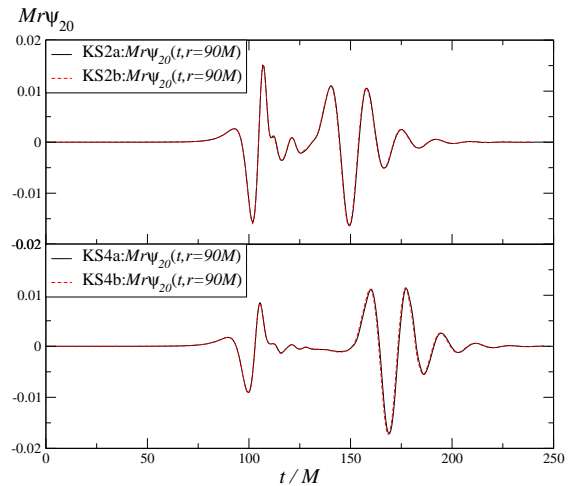}
  \caption{The $\ell=2$, $m=0$ multipole of $Mr\Psi_4$
           extracted at $r_{\rm ex}=90M$ for model
           KS2 (upper) and KS4 (lower panel) using
           gauge trajectories labeled in
           Table \ref{tab: bief_traj}
           as ``a'' (solid) and ``b'' (dashed curve).
           \label{fig: bief_traj}}
\end{figure}

We finally discuss the impact of the gauge trajectories
on the resulting waveforms. We have already mentioned in
Sec.~\ref{sec: gauge}
that the gauge trajectories need to closely resemble
the motion of the center of the apparent horizon
and are obtained iteratively by approximating the horizon
trajectory.
For this comparison we have constructed gauge trajectories
according to this procedure,
first, by fixing $v^z$ to be zero\footnote{This
velocity parameter is not
to be confused with that used in the calculation of the initial
data for $\gamma_{ij}$ and $K_{ij}$ which is always zero.}
and, second, by also adjusting this parameter in the
\begin{table}
  \caption{Parameters for the gauge trajectories used for the
           Kerr-Schild simulations.
           \label{tab: bief_traj}}
  \begin{ruledtabular}
  \begin{tabular}{l|ccccccc}
  Model         &  $z/M$     &  $v^z$     &  $a^z~M$        &  $j^z~M^2$        &  $q^z~M^3$      &  $\frac{t_1}{M}$  &  $\frac{t_2}{M}$  \\
  \hline
  KS1~a         &  $\pm5$    & 0          &  $\mp0.037$     &  $\pm0.0038$      &  0              &  10       &  35       \\
  KS1~b         &  $\pm5$    & $\mp0.08$  &  $\mp0.0061$    &  $\mp0.0002$      &  0              &  20       &  40       \\
  KS2~a         &  $\pm6$    & 0          &  $\mp0.029$     &  $\pm0.004$       &  $\mp0.000278$  &  25       &  50       \\
  KS2~b         &  $\pm6$    & $\mp0.06$  &  $\mp0.008$     &  $\pm0.0004$      &  $\mp 0.00002$  &  20       &  44       \\
  KS3~a         &  $\pm7$    & 0          &  $\mp0.022$     &  $\pm0.0027$      &  $\mp 0.000165$ &  25       &  57       \\
  KS3~b         &  $\pm7$    & $\mp0.04$  &  $\mp0.007$     &  $\pm0.0003$      &  $\mp 0.000012$ &  25       &  57       \\
  KS4~a         &  $\pm8$    & 0          &  $\mp0.018$     &  $\pm0.002$       &  $\pm0.000104$  &  34.5     &  84.7     \\
  KS4~b         &  $\pm8$    & $\mp0.03$  &  $\mp0.006$     &  $\pm0.00027$     &  $\mp 0.000012$ &  50       &  70       \\
  \end{tabular}
  \end{ruledtabular}
\end{table}
iterative procedure.
The parameters for the trajectories are listed in Table
\ref{tab: bief_traj} 
 and in Fig.~\ref{fig: bief_traj} we
show as examples the resulting waveforms extracted at
$r_{\rm ex}=90M$ for simulations
KS2a, b (upper panel) and KS4a, b (lower panel) obtained
using a resolution $h=1/24$.
The resulting waveforms are practically indistinguishable and the
differences in the energies for the initial pulse, the
merger signal and the total waveform are about $1~\%$
and thus significantly smaller than the uncertainties due
to the finite differencing and the extraction radius.
The same applies to variations from $2/3$ to zero
in the evolution parameter $\sigma$ in Eq.~\ref{eq: Gamma}.

\subsubsection{Misner data}

Finally, we estimate the numerical error of the evolutions
starting from Misner data. All evolutions of these data using the
Runge-Kutta time integration have resulted in significant contaminations
of the resulting waveforms by numerical noise. A comprehensive analysis
of the performance of different numerical schemes in evolutions
of Misner data and the underlying causes is beyond the scope of this work,
but we will show here that sufficiently accurate simulations
can be obtained by using the ICN scheme instead of Runge-Kutta
\begin{figure}
  \includegraphics[angle=-90,width=250pt]{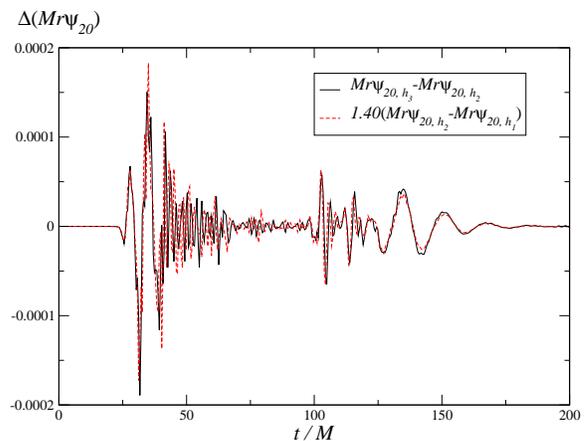}
  \caption{Convergence analysis of the $\ell=2$, $m=0$ mode
           of $\Psi_4$ obtained for model M4 using resolutions
           $h_1=1/400$, $h_2=1/360$ and $h_3=1/320$. The differences
           between the high resolution results has been amplified by
           a factor of $1.40$ expected for second-order convergence.
           \label{fig: new_conv_Misner}}
\end{figure}
and also choosing a small Courant factor of $1/8$.

The resulting differences in the $\ell=2$, $m=0$ mode of $Mr\Psi_4$
obtained for model M4
are shown in Fig.~\ref{fig: new_conv_Misner}. Compared with the
Brill-Lindquist and Kerr-Schild evolutions, we observe larger amounts
of high frequency noise in the early stages of the simulations
due to the spurious initial radiation. Still, the overall behavior
is compatible with the expected second-order
convergence. Using the same methods as before, we
find the resulting uncertainty
in the radiated energy due to finite differencing to be of the
order of $1~\%$. With regard to the extraction radius, we find
Misner data to behave similar to Brill-Lindquist data.
The resulting uncertainty due to the use of finite
radii is about $2~\%$ at $r_{\rm ex}=40M$ and $1~\%$ at $70M$ or $90M$.

\subsection{Results}

\begin{figure}
  \includegraphics[angle=-90,width=250pt]{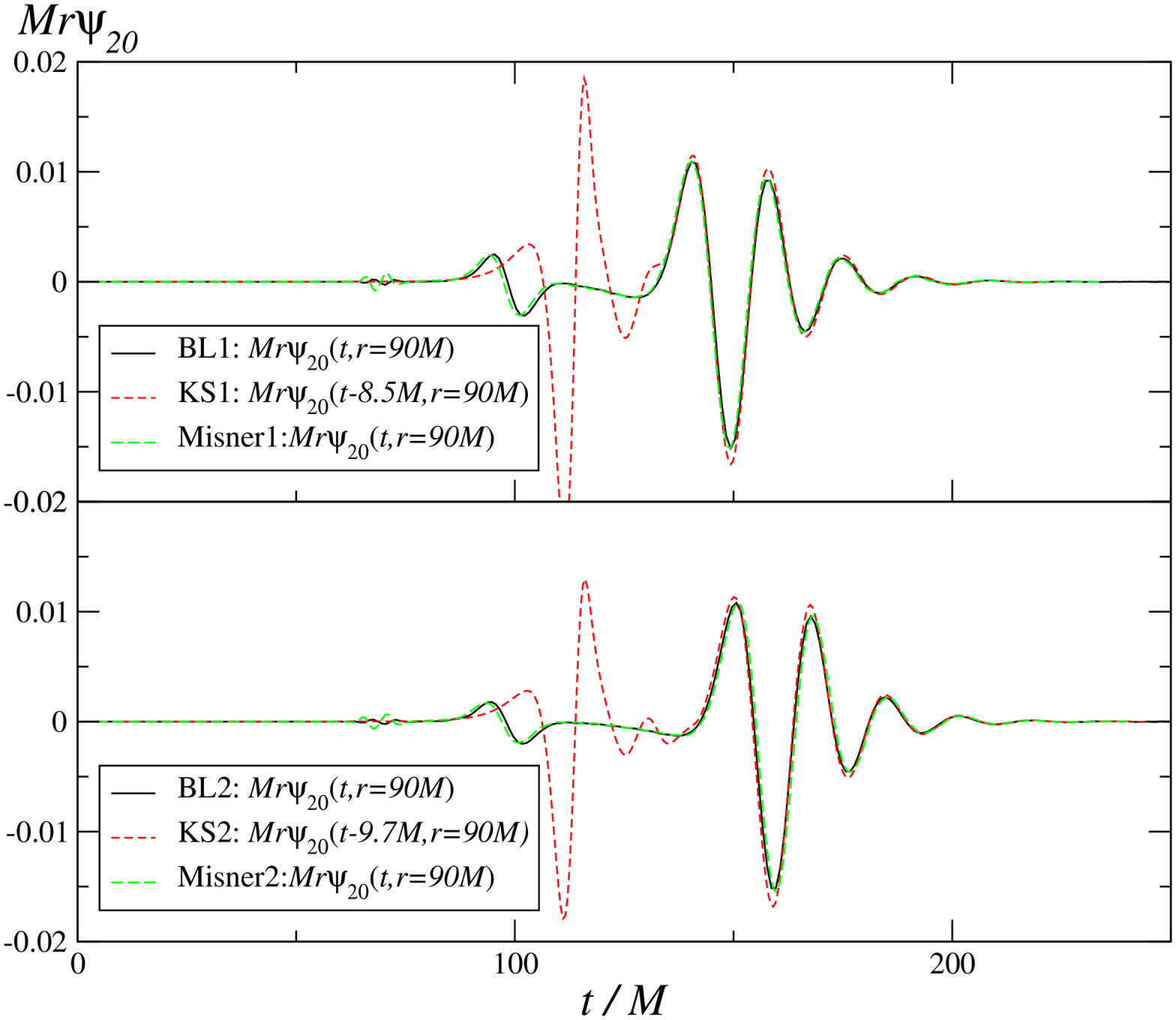}
  \includegraphics[angle=-90,width=250pt]{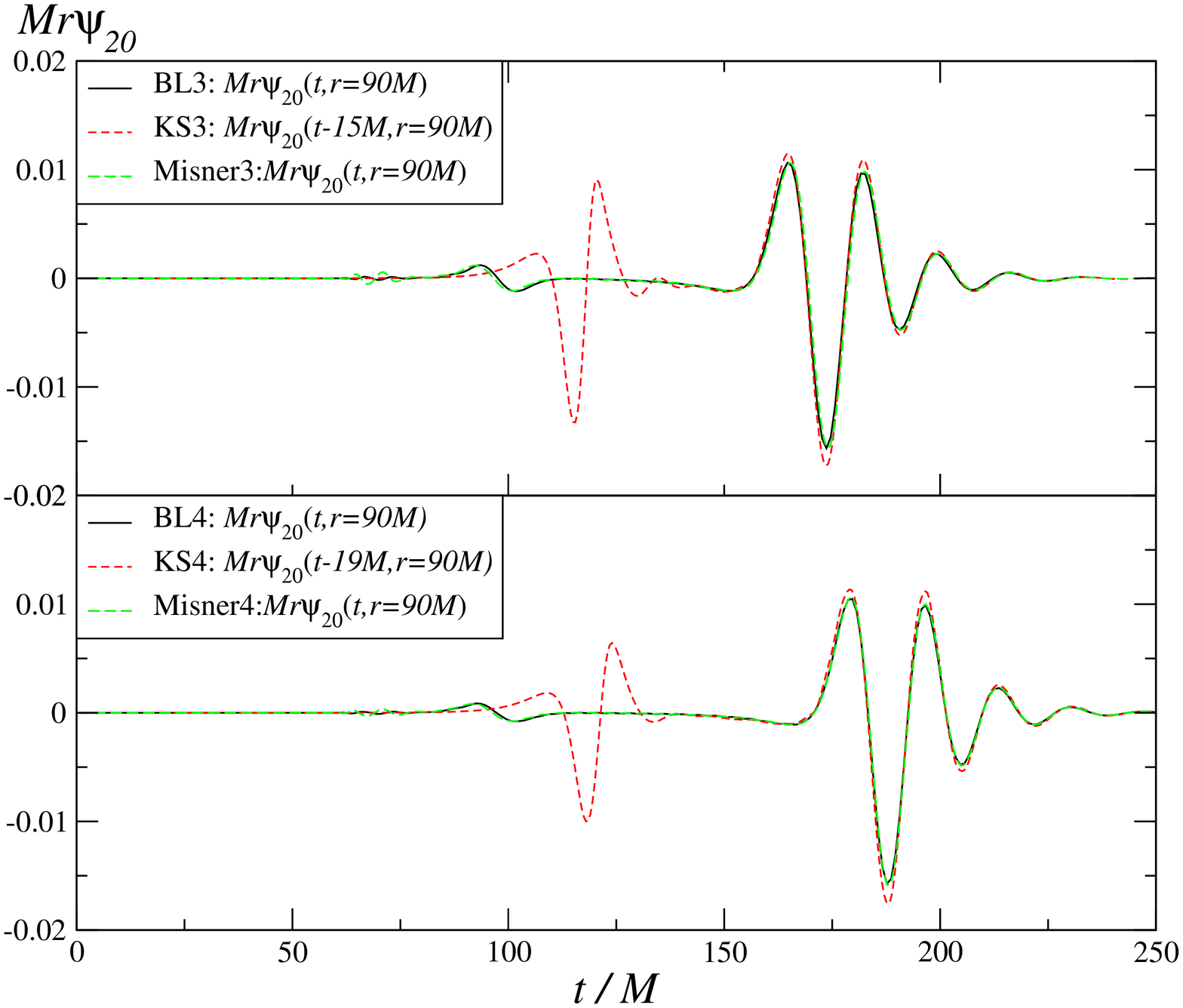}
  \caption{The $\ell=2$, $m=0$ mode
           of $Mr\Psi_4$ at $r_{\rm ex}=90M$
           for the Brill-Lindquist, Kerr-Schild and Misner
           versions
           of models 1--4. For presentation purposes, the
           Kerr-Schild data have been shifted in time to
           align the maxima of the waveforms.}
 \label{fig: bl_ks_mi}
\end{figure}

In order to compare the different initial-data types, we have evolved all
models listed in Table \ref{tab: models} with the grid setups
described in Table \ref{tab: grids}. We summarize the results in
Fig.~\ref{fig: bl_ks_mi} which shows the $\ell=2$, $m=0$ modes
obtained at extraction radius $r_{\rm ex}=90M$ and
Table \ref{tab: models} where we list the initial parameters
and the total radiated energy $E_{\rm tot}$ as well as the contributions
of the spurious initial pulse $E_{\rm ini}$ and the merger signal
$E_{\rm rad}$. The radiated energies have been extracted
at $r_{\rm ex}=90M$ and for Kerr-Schild data
we also give the extrapolated values for $r_{\rm ex}\rightarrow \infty$.
The uncertainties are those obtained
in the previous section. In the case of Kerr-Schild data
starting from small separations,
the error for $E_{\rm ini}$ and $E_{\rm rad}$ is
amplified significantly
by the uncertainties in separating the initial pulse
from the merger signal.

In Fig.~\ref{fig: bl_ks_mi} we note the substantially larger
amount of artificial radiation due to the initial data in the
Kerr-Schild case (dashed curve). Second, we observe excellent
agreement between the waveforms obtained from Misner and
Brill-Lindquist data (the solid and long-dashed curves are
practically indistinguishable) and
good qualitative agreement with the Kerr-Schild results.
There remains, however, a small systematic deviation to the
effect that the Kerr-Schild waves have a $5-10~\%$
larger amplitude. We have already seen, however, that the
finite extraction radius results in an overestimation of the
wave amplitudes, in particular in the Kerr-Schild case.

In order to quantify this effect, we consider the radiated
energies in Table \ref{tab: models}. We first note that
within the uncertainties
Brill-Lindquist and Misner data result in identical amounts
of energy in the initial-data pulse, the merger
waveform as well as the total radiation.
In contrast, the total radiated energy obtained from the Kerr-Schild
data is significantly larger than that of Misner and
Brill-Lindquist simulations. This excess energy is largely due
to the spurious initial pulse, however, and for models KS1 and KS2,
the energy contained in the physically important merger waveform
agrees within the error bounds with its Misner and Brill-Lindquist
counterparts. In particular, this is true for the values obtained
from extrapolation to infinite extraction radius. The situation
becomes more complicated, however, for the Kerr-Schild models starting
from larger separations, in particular for model KS4. With the
error estimates obtained in the previous section, we obtain a lower
limit of $E_{\rm rad}=0.0580~\%$ of the ADM mass which exceeds the
upper limit of simulation BL4 of $E_{\rm rad}=0.0563$ by about $3~\%$.
While we have taken into account in the derivation of these results
the errors arising from finite differencing and the wave extraction
at finite radius, it is possible that systematic errors are responsible
for the remaining discrepancy. We have seen in Sec.~\ref{sec: test_bl}
that small boosts give rise to radiated energies a few per cent
larger than those obtained for initially time symmetric data.
A further systematic error results from the constraint violations
inherent to the superposed Kerr-Schild data.
Unfortunately there
exists, to our knowledge,
no literature on time evolutions of the constraint-solved
version of the Kerr-Schild data. Filling this gap is beyond the
scope of this paper as it requires the addition of elliptic solvers,
currently not available in the {\sc Lean} code. It is therefore currently not
possible to rigorously quantify the impact of the constraint violations
on the resulting waveforms. We note, however, that the amount of
spurious initial gravitational radiation inherent to the superposed Kerr-Schild
data is significantly larger than the discrepancies we
observe. If this spurious initial radiation is a signature of the
constraint violations, it is certainly possible that the discrepancies
observed here are due to the constraint violations of the Kerr-Schild data.

A further interesting question is the dependence of the radiated energies
on the initial black-hole separations. We have already noticed that the
amount of spurious initial radiation decreases at larger separations
as is expected. Correspondingly, we observe the expected
increase in the energy radiated in the infall and merger
at larger initial separations. This increase is relatively weak, though,
for the cases studied here, especially for Brill-Lindquist and Misner data.
It would be desirable to
probe a larger range of initial distances, in particular smaller
separations, to study this behavior in more detail.
Unfortunately, such an extension encounters difficulties at
either end of the spectrum.

In the case of Kerr-Schild data,
separations smaller than that of KS1 lead to a severe contamination
of the actual signal by the spurious wave content and thus do not
allow a physically meaningful interpretation.
At the upper end, we are limited by the
construction of suitable gauge trajectories.
The prolonged infall puts stronger demands on the fine tuning of
the gauge trajectories. So far, we have not managed to obtain
stable evolutions starting from Kerr-Schild data with $D>16~M$.

In the case of Brill-Lindquist data, we do not encounter such difficulties
with the gauge
because of the universality of the live-gauge
conditions (\ref{eq: alpha})--(\ref{eq: B}).
Results starting from small separations, however, are subject to
the difficulties due to the initial wave burst.
This is illustrated by evolving a set of
\begin{figure}
  \includegraphics[angle=-90,width=250pt]{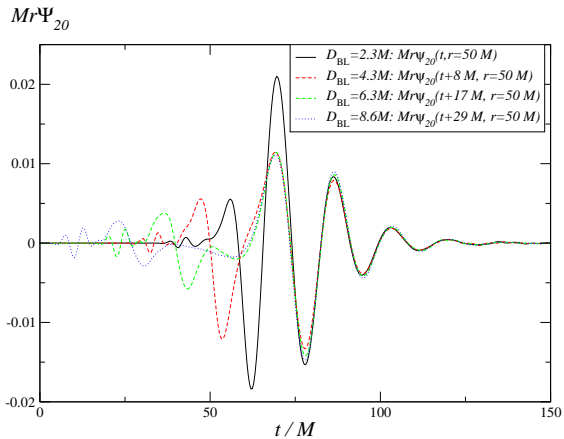}
  \caption{Waveforms obtained from Brill-Lindquist data starting
           at various initial separations. The data have been shifted
           in time to align the maxima of the waveforms.}
 \label{fig: new_bl_D}
\end{figure}
Brill-Lindquist data starting with initial separations
$D_{\rm BL}=2.3\,M$, $4.3\,M$ and $6.3\,M$. The resulting waveforms
as obtained with the setup labelled ``ISCO'' in Table \ref{tab: grids}
are shown together with that of model BL1 in Fig.\,\ref{fig: new_bl_D}.
We clearly notice a substantial contamination of the waveforms at small
separations by radiation inherent to the initial data.
In consequence, an accurate calculation
of the gravitational wave energy generated in the head-on
collision becomes highly nontrivial even for
Brill-Lindquist data.
For this reason the comparison performed in this study is currently
limited to the window of binding energies or separations covered by
the results in Table \ref{tab: models}.

\section{Summary and conclusions}
\label{sec: conclusions}

In this work, we have presented in detail a numerical code designed
for the simulation of black-hole binaries in the framework of
three-dimensional, nonlinear general relativity. The code facilitates
black-hole evolutions using different initial-data types and evolution
techniques.

It has been
demonstrated that the code is capable of evolving state-of-the-art
binary-black-hole orbits using the recently developed moving-puncture
technique. With regard to the accuracy of the results, we find
it crucial to use a fourth-order discretization of the spatial
derivatives appearing in the BSSN formulation of the Einstein
field equations.
The resulting simulations
yield convergent waveforms which agree well with
results presented in the literature. The same holds for the radiated
energy which we estimate to be $3.408\%\pm 0.102\%$ of the total ADM mass.
The code is thus suitable
for detailed studies of various types of multiple black-hole
simulation with regard to the generation of accurate
waveform templates.

In preparation for the comparison of black-hole collisions using different
types of initial data, we test the code's performance
and estimate in detail the error margins associated with
the different evolutions. Specifically, we
separately demonstrate convergence of the code for
simulations starting from Brill-Lindquist, superposed Kerr-Schild
and Misner data. We also study in depth the dependence of
the resulting waveforms on the extraction radii. While the
resulting uncertainties are relatively small for Brill-Lindquist
and Misner data, we find the use of finite extraction radii
to be the dominant error source for simulations of Kerr-Schild data.
In the case of Brill-Lindquist data we further demonstrate the
codes reliability by comparing head-on collisions with results
available in the literature.

We use the code to provide a detailed comparison of
black-hole-binary head-on collisions using
all three data types.
In addition to the total mass of the system,
either initial configuration has one free parameter
which is specified by fixing the binding energy $E_{\rm b}/M$ of
the system. We have compared the resulting
waveforms for four initial configurations.

The resulting waveforms obtained from Brill-Lindquist and
Misner data
show excellent agreement and predicts an energy
radiated in the infall and merger of
about $0.052\%-0.056\%$ of the total ADM mass $M$ with the
exact value increasing with the initial black-hole separation.
In parallel the amount of energy due to spurious radiation inherent
to the conformally flat initial data
decreases from $0.0031~\%~M$
to $0.0009~\%~M$ as the initial separation
of the holes is increased from a binding energy $E_{\rm b}=-0.029~M$
to $-0.0169~M$. The uncertainties in these results for both data
types are of the order of a few percent.
While this good agreement might be expected, given the similar
nature of two data types, it is reassuring to confirm this
expectation with high accuracy using current numerical techniques.

In the case of the superposed Kerr-Schild data, we observe
a substantially larger amount of energy in the
spurious gravitational radiation due to the initial data,
about a factor 15 larger than for the other data types.
For sufficiently large separations,
most of this spurious
wave content radiates away before the merger of the
holes and can thus be distinguished from the actual signal of the
merger and ring-down. For the smaller separations studied, however,
this distinction becomes more difficult and leads to a
non-negligible uncertainty in the amount of energy radiated
in the infall and merger of the holes.

In comparison with the conformally flat data types,
the merger waveforms extracted at finite radius
show larger amplitudes
in the case of the Kerr-Schild simulations by $5-10~\%$.
The agreement of the resulting radiated energies in the
infall and merger becomes much better, though, after
extrapolating results to infinite extraction radius.
Still, for large black-hole separations
there remains a discrepancy of a few $\%$ in the
merger energy between Kerr-Schild data on the one side and
Brill-Lindquist and Misner data on the other, even when
taking into account remaining uncertainties in the simulations.
We can therefore not
rule out systematic errors affecting the accuracy of the
Kerr-Schild simulations.

Such systematic errors can arise from the fact that the initial
data are not inherently time symmetric and might imply small initial
boosts of the individual holes.
By evolving puncture data with
nonvanishing Bowen-York momentum we have shown that
small boosts can account for the discrepancies of the observed
magnitude.
A second systematic error arises from the constraint violations
of the Kerr-Schild data. In particular, we observe
the energy contained in the
spurious initial radiation, likely to be a
signature of the constraint violations, to be significantly larger
than the differences we observe.

Finally, we mention future directions of research encouraged
by this study. First, it will be valuable to understand
the origin for the relatively large uncertainties
in the wave amplitudes obtained from Kerr-Schild data at
finite extraction radius or, conversely, why wave extraction
appears to work remarkably well for the Misner and puncture
simulations at radii significantly smaller than the wave zone.
Second, it will be important to produce evolutions
of the constraint-solved version of the Kerr-Schild data
and compare the results with those of the present study.
An important question in this regard also concerns the
amount of spurious initial radiation. A key advantage of
Kerr-Schild--type initial data over conformally flat data
is the fact that they contain the Kerr solution as a limit.
They are therefore particularly promising candidates for
producing initial data sets containing black holes
with very large spins with minimal artificial radiation
(see, for example,
\cite{Hannam2007a} for a discussion of artificial
radiation in black-hole spacetimes with large spins).
It is thus important to study how the solving of
the constraints reduces the large amounts of artificial
radiation observed here in the nonspinning case.

\begin{acknowledgments}
I thank Marcus Ansorg, Erik Schnetter and Jonathan Thornburg for
providing The {\sc TwoPuncture} thorn, {\sc Carpet} and {\sc AHFinderDirect}.
I further thank Bernd Br\"ugmann, Jose Gonzalez, Mark Hannam, Sascha Husa
and Christian K\"onigsd\"orffer
for illuminating discussions in all areas of this work.
I also thank Bernard Kelly, Pablo Laguna, Ken Smith, Deirdre
Shoemaker and Carlos Sopuerta
for fruitful discussions concerning all aspects of the Kerr-Schild evolutions.
This work was supported by the DFG grant SFB/Transregio~7
``Gravitational Wave Astronomy'', and computer time allocations at
HLRS Stuttgart and LRZ Munich. I acknowledge support from the ILIAS
Sixth Framework programme.
\end{acknowledgments}

\appendix

\section{The BSSN evolution equations}
\label{sec: BSSN}

The BSSN equations are implemented in the {\sc Lean code} using either
the set of variables defined in Eq.~(\ref{eq: BSSN_vars}) or using the
variable $\chi$ in place of $\phi$ as defined in Eq.~(\ref{eq: BSSN_chi}).
The $\phi$ version of the BSSN equations used in {\sc Lean} is given by
\begin{widetext}
\begin{eqnarray}
  \partial_t \tilde{\gamma}_{ij} &=& \beta^m \partial_m \tilde{\gamma}_{ij}
        + 2\tilde{\gamma}_{m(i} \partial_{j)} \beta^m - \frac{2}{3}
        \tilde{\gamma}_{ij} \partial_m \beta^m -2\alpha \tilde{A}_{ij},
        \label{eq: gamma} \\
  \partial_t \phi &=& \beta^m \partial_m \phi + \frac{1}{6} (\partial_m \beta^m
        - \alpha K), \\
  \partial_t \tilde{A}_{ij} &=& \beta^m \partial_m \tilde{A}_{ij}
        + 2\tilde{A}_{m(i} \partial_{j)} \beta^m - \frac{2}{3} \tilde{A}_{ij}
        \partial_m \beta^m + e^{-4\phi} \left( \alpha R_{ij}
        - D_i D_j \alpha\right)^{\rm TF} + \alpha \left( K\,\tilde{A}_{ij}
        - 2 \tilde{A}_i{}^m \tilde{A}_{mj} \right), \\
  \partial_t K &=& \beta^m \partial_m K - D^m D_m \alpha + \alpha \left(
        \tilde{A}^{mn} \tilde{A}_{mn} + \frac{1}{3} K^2 \right),
        \label{eq: tracek} \\
  \partial_t \tilde{\Gamma}^i &=& \beta^m \partial_m \tilde{\Gamma}^i
        - \tilde{\Gamma}^m \partial_m \beta^i
        + \frac{2}{3} \tilde{\Gamma}^i \partial_m \beta^m
        + 2 \alpha \tilde{\Gamma}^i_{mn} \tilde{A}^{mn}
        + \frac{1}{3} \tilde{\gamma}^{im}\partial_m \partial_n \beta^n
        + \tilde{\gamma}^{mn} \partial_m \partial_n \beta^i \nonumber\\
     && - \frac{4}{3} \alpha \tilde{\gamma}^{im} \partial_m K
        + 2\tilde{A}^{im} \left( 6 \alpha \partial_m \phi
          - \partial_m \alpha \right)
        -\left( \sigma + \frac{2}{3}\right) \left(\tilde{\Gamma}^i
          -\tilde{\gamma}^{mn}\tilde{\Gamma}^i_{mn} \right)
          \partial_k \beta^k.
        \label{eq: Gamma}
\end{eqnarray}
\end{widetext}
Here $D_i$ is the covariant derivative operator
and $R_{ij}$ the Ricci tensor associated
with the physical three-metric $\gamma_{ij}$ and
the superscript ${}^{\rm TF}$ denotes the trace-free part.
We also note that the last term in the evolution equation (\ref{eq: Gamma})
vanishes in the continuum limit by virtue of the
definition of $\tilde{\Gamma}^i$ in Eq.\,(\ref{eq: BSSN_vars}). With the
addition of this term we follow Yo et.\,al.\,\cite{Yo2002} who
introduced this
modification to improve the stability of the BSSN formulation in cases
of relaxed symmetry assumptions of the spacetime under study. We set
the free parameter $\sigma$ in this term to $2/3$.

The $\chi$ version of the evolution system is obtained by substituting
$\chi=e^{-4\phi}$. The evolution equations for the variables $\chi$,
$\tilde{A}^{ij}$ and $\tilde{\Gamma}^i$ are then given by
\begin{widetext}
\begin{eqnarray}
  \partial_t \chi &=& \beta^m \partial_m \chi + \frac{2}{3}
      (\alpha K - \partial_m \beta^m),
      \label{eq: chi} \\
  \partial_t \tilde{A}_{ij} &=& \beta^m \partial_m \tilde{A}_{ij}
        + 2\tilde{A}_{m(i} \partial_{j)} \beta^m - \frac{2}{3} \tilde{A}_{ij}
        \partial_m \beta^m + \chi \left( \alpha R_{ij}
        - D_i D_j \alpha\right)^{\rm TF} + \alpha \left( K\,\tilde{A}_{ij}
        - 2 \tilde{A}_i{}^m \tilde{A}_{mj} \right), \\
  \partial_t \tilde{\Gamma}^i &=& \beta^m \partial_m \tilde{\Gamma}^i
        - \tilde{\Gamma}^m \partial_m \beta^i
        + \frac{2}{3} \tilde{\Gamma}^i \partial_m \beta^m
        + 2 \alpha \tilde{\Gamma}^i_{mn} \tilde{A}^{mn}
        + \frac{1}{3} \tilde{\gamma}^{im}\partial_m \partial_n \beta^n
        + \tilde{\gamma}^{mn} \partial_m \partial_n \beta^i \nonumber \\
     && - \frac{4}{3} \alpha \tilde{\gamma}^{im} \partial_m K
        - \tilde{A}^{im} \left( 3 \alpha \frac{\partial_m \chi}{\chi}
          + 2\partial_m \alpha \right)
        -\left( \sigma + \frac{2}{3}\right) \left(\tilde{\Gamma}^i
          -\tilde{\gamma}^{mn}\tilde{\Gamma}^i_{mn} \right)
          \partial_k \beta^k,
        \label{eq: gamma2}
\end{eqnarray}
\end{widetext}
while Eqs.~(\ref{eq: gamma}) and (\ref{eq: tracek}) remain valid without
modification.

\section{The ADM variables of a single boosted Kerr-Schild black hole}
\label{sec: bKS}

Purpose of this section is to calculate the ADM functions $\gamma_{ij}$,
$K_{ij}$, $\alpha$ and $\beta^i$ as functions of the laboratory coordinates
$x^{\mu}$ for a nonspinning
boosted black hole with mass parameter $m$
in Kerr-Schild coordinates moving with
speed $v^i$ in the laboratory rest frame\footnote{See, for example,
\cite{Moreno2002} for a discussion of the superposition of Kerr-Schild
holes with nonvanishing spin.}.
The rest frame coordinates
of the black hole are related to the laboratory coordinates
by a Lorentz transformation
\begin{equation}
  x^{\bar{\alpha}} = \Lambda^{\bar{\alpha}}{}_{\mu} x^{\mu},
  \label{eq: coordinates}
\end{equation}
where the transformation matrix is given by
\begin{equation}
  \Lambda^{\bar{\alpha}}{}_{\mu} = \left(
      \begin{array}{cc}
         \gamma       & -\gamma v_m \\
         -\gamma v^a\quad  & \delta^a{}_m + (\gamma-1) \frac{v^a v_m}{\vec{v}^2}
      \end{array}
      \right).
\end{equation}
In the black-hole rest frame, the spacetime metric is given by (see
e.~g.~\cite{Matzner1998})
\begin{equation}
  g_{\bar{\alpha} \bar{\beta}} = \eta_{\bar{\alpha} \bar{\beta}}
       + 2H\ell_{\bar{\alpha}} \ell_{\bar{\beta}},
\end{equation}
where
\begin{eqnarray}
  H &=& \frac{m}{\bar{r}},
      \label{eq: bKS_H} \\
  \ell_{\bar{\alpha}} &=& \left[ 1,\,\,\frac{x_{\bar{a}}}{\bar{r}} \right],
  \label{eq: bKS_ell} \\
  \bar{r} &=& x^{\bar{a}} x_{\bar{a}},
\end{eqnarray}
and indices of $x^{\bar{a}}$ and $v^{\bar{a}}$
are raised and lowered with the flat space
metric $\delta_{\bar{a}\bar{b}}$.

The spacetime metric in the laboratory frame is obtained from that
in the black-hole frame by a Lorentz transformation
\begin{equation}
  g_{\mu \nu} = \Lambda^{\bar{\alpha}}{}_{\mu} \Lambda^{\bar{\beta}}{}_{\nu}
                g_{\bar{\alpha} \bar{\beta}}
              = \eta_{\mu \nu} + 2H \ell_{\mu} \ell_{\nu},
  \label{eq: bKS_metric}
\end{equation}
where we have used the fact that $H$ and $\ell_{\mu}$ behave like a scalar and
vector, respectively, and the Minkowski metric is invariant
under Lorentz transformations.

From the spacetime metric we directly obtain the 3-metric, its inverse
as well as lapse and shift
\begin{eqnarray}
  \gamma_{mn} &=& \delta_{mn} + 2H \ell_m \ell_n,
      \label{eq: bKS_3metric} \\
  \gamma^{mn} &=& \delta^{mn} - 2H \delta^{mk} \delta^{nl} \ell_k \ell_l
                  / [1+2H(\ell_0)^2], \\
  \alpha &=& \left[1+2H(\ell_0)^2\right]^{-1/2},
      \label{eq: bKS_lapse} \\
  \beta_m &=& 2H\ell_0 \ell_m, \\
  \beta^m &=& 2H\ell_0 \delta^{mk}\ell_k / [1+2H(\ell_0)^2].
      \label{eq: bKS_shift}
\end{eqnarray}
The extrinsic curvature is obtained from the derivatives of the three-metric
according to
\begin{eqnarray}
  K_{mn} &=& -\frac{1}{2\alpha} \left( \partial_t \gamma_{mn}
             -\mathcal{L}_{\beta} \gamma_{mn} \right) \nonumber \\
         &=& -\frac{1}{2\alpha} \left( \partial_t \gamma_{mn}
             - 2D_{(m}\beta_{n)} \right),
  \label{eq: bKS_K}
\end{eqnarray}
where $\mathcal{L}_{\beta}$ denotes the Lie derivative along the shift vector
and $D_m$ the three-dimensional covariant derivative operator.

The derivatives of the three-metric are most conveniently expressed in terms
of $H$ and $\ell_m$
\begin{eqnarray}
  \partial_t \gamma_{mn} &=& 2(\ell_m \ell_n \partial_t H + H\ell_n
        \partial_t \ell_m + H \ell_m \partial_t \ell_n), \\
  D_{(m}\beta_{n)} &=& 2\left[ \ell_0 \ell_{(m}\partial_{n)}H
        + H\ell_{(m}\partial_{n)} \ell_0 + H\ell_0
        \partial_{(m}\ell_{n)} \right] \nonumber \\
       && - \Gamma^{k}_{mn}\beta_k.
\end{eqnarray}
Finally, the derivatives of $H$ and $\ell_{\mu}$ are given by
\begin{eqnarray}
  \partial_{\mu} H &=& -\Lambda^{\bar{\alpha}}{}_{\mu} \frac{mx^{\bar{\alpha}}}
        {\bar{r}^3}, \\
  \partial_{\mu} \ell_{\nu} &=& \Lambda^{\bar{a}}{}_{\mu}
        \Lambda^{\bar{b}}{}_{\nu} \delta_{\bar{a}\bar{b}} \frac{1}{\bar{r}}
        -\Lambda^{\bar{a}}{}_{\mu}
        \Lambda^{\bar{b}}{}_{\nu} \frac{x_{\bar{a}} x_{\bar{b}}}{\bar{r}^3}.
\end{eqnarray}
In summary, the function for calculating the ADM variables of a boosted
black hole in Kerr-Schild coordinates at a particular point
requires as input the coordinates of the point in the laboratory frame
as well as the velocity $\vec{v}$ of the hole. First, the coordinates
are transformed into the rest frame of the black hole according to
Eq.~(\ref{eq: coordinates}). Next $H$ and $\ell_{\bar{\alpha}}$ follow
from Eqs.~(\ref{eq: bKS_H}), (\ref{eq: bKS_ell}) and give the spacetime
metric components (\ref{eq: bKS_metric}). The three-metric, lapse and
shift follow from Eqs.~(\ref{eq: bKS_3metric}), (\ref{eq: bKS_lapse})
and (\ref{eq: bKS_shift}). Together with the extrinsic curvature
(\ref{eq: bKS_K}), these are returned to the calling function as
the ADM variables of the boosted Kerr-Schild metric.

\section{The electromagnetic decomposition of the Weyl tensor and wave extraction}
\label{sec: wave_extraction}
We calculate the Newman-Penrose scalar $\Psi_4$ from the Weyl tensor
via
\begin{equation}
  \Psi_4 = C_{\alpha \beta \gamma \delta} n^{\alpha} \bar{m}^{\beta}
       n^{\gamma} \bar{m}^{\delta},
  \label{eq: Psi4}
\end{equation}
where $n$ and $\bar{m}$ form part of a null-tetrad $\ell$, $n$, $m$, $\bar{m}$
such that all their inner products vanish except
\begin{equation}
  -\ell\cdot n = 1 = m\cdot \bar{m}
\end{equation}
Specifically, we construct $\ell$, $n$ and $m$ from the orthonormal triad
vectors $u$, $v$ and $w$ according to
\begin{eqnarray}
  \ell^{\alpha} &=& \frac{1}{\sqrt{2}} \left( \hat{n}^{\alpha}
      + u^{\alpha} \right), \nonumber \\
  n^{\alpha} &=& \frac{1}{\sqrt{2}} \left( \hat{n}^{\alpha}
      - u^{\alpha} \right), \nonumber \\
  m^{\alpha} &=& \frac{1}{\sqrt{2}} \left( v^{\alpha} + iw^{\alpha} \right),
  \label{eq: tetrad}
\end{eqnarray}
where $\hat{n}^{\mu}$ is the timelike orthonormal vector. The triad
$u$, $v$, $w$ is constructed via Gram-Schmidt orthonormalization
starting with
\begin{eqnarray}
  u^i &=& \left[ x\,\,,y\,\,,z \right], \nonumber \\
  v^i &=& \left[ xz\,\,,yz\,\,,-x^2-y^2\right], \nonumber \\
  w^i &=& \epsilon^i{}_{mn}v^m w^n,
\end{eqnarray}
where $\epsilon^{imn}$ represents the three-dimensional Levi-Civita tensor.

In the decomposition of the Weyl tensor we follow
the presentation of Friedrich \cite{Friedrich1996}. The electric and magnetic
part of the gravitational field are given by
\begin{eqnarray}
  E_{\alpha \beta} &=& \bot^{\mu}{}_{\alpha} \bot^{\nu}{}_{\beta}
       C_{\mu \rho \nu \sigma} n^{\rho}  n^{\sigma}, \\
  B_{\alpha \beta} &=& \bot^{\mu}{}_{\alpha} \bot^{\nu}{}_{\beta}
       {}^{*}C_{\mu \rho \nu \sigma},
\end{eqnarray}
where $\bot^{\mu}{}_{\alpha} =\delta^{\mu}{}_{\alpha}
+\hat{n}^{\mu}\hat{n}_{\alpha}$ is the projector onto the
spacelike hypersurface and the ${}^{*}$ denotes the Hodge dual.
By virtue of the Gauss-Codazzi equations (see, e.~g.~\cite{Gourgoulhon2007}),
one can express the electromagnetic parts in terms of ``3+1'' variables
according to
\begin{eqnarray}
  E_{ij} &=& R_{ij} - \gamma^{mn}\left( K_{ij} K_{mn} - K_{im}K_{jn} \right),
      \nonumber\\
  B_{ij} &=& \gamma_{ik}\epsilon^{kmn}D_m K_{nj}.
  \label{eq: elm3+1}
\end{eqnarray}
The Weyl tensor is then given in terms of the electric and magnetic part
by Eq.~(3.10) of Ref.~\cite{Friedrich1996}. Inserting this relation
together with Eqs.~(\ref{eq: tetrad}), (\ref{eq: elm3+1}) into the definition
(\ref{eq: Psi4}) enables us to express $\Psi_4$ exclusively in terms of
``3+1'' quantities
\begin{eqnarray}
  \Psi_4 = \frac{1}{2} \left[ E_{mn}(v^m v^n - w^m w^n)
                            - B_{mn}(v^m w^n + w^m v^n) \right]
           \nonumber \\
         - \frac{i}{2} \left[ E_{mn}(v^m w^n - w^m v^n)
                            + B_{mn}(w^m w^n + v^m v^n) \right].
           \nonumber
\end{eqnarray}
In practice, $\Psi_4$ is calculated using this relation on the entire
Cartesian grid and then interpolated onto coordinate spheres of
different extraction radii. We then apply a mode decomposition
using spherical harmonics $Y^{-2}_{\ell m}$
of spin-weight $-2$ (cf.~Eq.~(42) in
Ref.~\cite{Bruegmann2006a}) according to
\begin{eqnarray}
  \Psi_4(t,\theta,\phi) &=& \sum_{\ell,m} \psi_{\ell m}(t) Y^{-2}_{\ell m}
          (\theta,\phi), \\
  \psi_{\ell m}(t) &=& \int \Psi_4(t,\theta,\phi)\overline{Y^{-2}_{\ell m}}(\theta, \phi)
      d\Omega.
\end{eqnarray}
In this context we note that $\Psi_4$ is always extracted onto the entire
coordinate sphere $\theta=0\ldots \pi$, $\phi=0\ldots 2\pi$, even when
underlying symmetry of the physical problem is used to reduce the
computational domain to a bitant or octant. In those cases we use the
fact that the real part of $\Psi_4$ behaves like a scalar while the
imaginary part of $\Psi_4$ behaves like a pseudo scalar, i.~e.~reverses
its sign across symmetry boundaries.

\section{Performance of the code}
\label{sec: performance}

The majority of simulations presented in this work have been performed
using a 24 node Linux cluster. Each node contains four AMD 2200 GHz processors
and provides 8 Gb of memory. Parallelization is implemented using
the $\tt MPICH$ version 1.2.6 2004/08/08
libraries. The code has been compiled with version 4.0.2 2005091 of
the $\tt gcc$, $\tt g++$ and $\tt gfortran$ compilers. Compared with alternative
architectures (cf.~Ref.~\cite{Marronetti2007}), we have noticed that this
\begin{table}
  \caption{Performance summary for representative simulations presented
           in this work.
           \label{tab: performance}}
  \begin{ruledtabular}
  \begin{tabular}{l|cccc}
  Simulation    &  $dt/dx$  &  \#CPU   &  Memory        &  Speed \\
                &           &          &  [Gb]          &  [M/CPUh] \\
  \hline
  R1 $(h=1/40)$ &  1/2      &  12      &  20.1          &  0.620  \\
  R1 $(h=1/44)$ &  1/2      &  16      &  26.4          &  0.347  \\
  R1 $(h=1/48)$ &  1/2      &  24      &  35.2          &  0.225  \\
  \hline
  BL4 $(h=1/40)$&  1/2      &   8      &  9.9           &  0.854  \\
  BL4 $(h=1/44)$&  1/2      &  12      &  12.5          &  0.582  \\
  BL4 $(h=1/48)$&  1/2      &  16      &  18.2          &  0.350  \\
  \hline
  KS4 $(h=1/20)$&  1/4      &   8      &  8.1           &  0.571  \\
  KS4 $(h=1/24)$&  1/4      &   8      &  14.0          &  0.315  \\
  KS4 $(h=1/28)$&  1/4      &  16      &  22.5          &  0.171  \\
  \hline
  M4 $(h=1/320)$&  1/8      &   8      &  9.1           &  0.269  \\
  M4 $(h=1/360)$&  1/8      &  12      & 12.3           &  0.162  \\
  M4 $(h=1/400)$&  1/8      &  12      & 17.5           &  0.111  \\
  \end{tabular}
  \end{ruledtabular}
\end{table}
architecture requires about $25~\%$ more memory resources for identical
simulations.

In Table \ref{tab: performance} we summarize the performance of the
code for simulations R1, BL4, KS4 and M4 of Table \ref{tab: models}
using this architecture. The columns show the Courant
factor $dt/dx$ which scales linearly with the code's speed,
the number of processors used in the simulation, the
required memory as well as the speed. The latter
is measured in physical time in units of the ADM mass
$M$ of the system per real time and processor.

Regarding the memory usage we observe minor variations, typically
below $5~\%$, in the course of the simulation. This is due to the
merger of refinement components as the black holes approach
each other. The merger of refinement components also leads to
an increase in speed because the costly regridding operation is no
longer required and the total number of grid points of the
merged refinement component is smaller than the sum of the two
individual ones prior to merging.
All reported speeds are averages over the entire simulation.


\end{document}